\documentclass[nofootinbib, reprint,amsmath, superscriptaddress, amssymb, aps, prb]{revtex4-2}
\usepackage{lipsum}  
\usepackage{stackengine}
\usepackage{graphicx}
\usepackage[colorlinks=true, citecolor=blue, urlcolor=blue, linkcolor=blue, bookmarks=false, hypertexnames=true]{hyperref}
\usepackage{amsmath}
\usepackage{soul}
\usepackage{dcolumn}
\usepackage{lipsum}  
\usepackage{float}
\usepackage{comment}
\newcommand\thefontsize{The current font size is: \f@size pt}
\usepackage{bm}
\usepackage{color}
\usepackage{xcolor}
\usepackage[normalem]{ulem}
\usepackage{lineno}
\begin{document}
\renewcommand{\thesection}{\Roman{section}}

\preprint{APS/123-QED}

%%%%%%%%%%%%%%%%%%%%%%%%%%%%%%%%%%%%%%%%%%%%%%%%%%%%%%%%%%%%%%%%
\title{Impact of exciton-phonon and exciton-magnon interaction on transport in anisotropic 2D magnetic semiconductors}
%\date{\today}
%%%%%%%%%%%%%%%%%%%%%%%%%%%%%%%%%%%%%%%%%%%%%%%%%%%%%%%%%%%%%%%%

\author{Daniel Erkensten}
\email{Email: daniel.erkensten@physik.uni-marburg.de}
\affiliation{Department of Physics, Philipps-Universit{\"a}t Marburg, 35037 Marburg, Germany}
\affiliation{mar.quest|Marburg
Center for Quantum Materials and Sustainable Technologies,
35032 Marburg, Germany}

\author{Katarzyna Sadecka}
\affiliation{Department of Physics, Philipps-Universit{\"a}t Marburg, 35037 Marburg, Germany}
\affiliation{mar.quest|Marburg
Center for Quantum Materials and Sustainable Technologies,
35032 Marburg, Germany}
\author{Marie-Christin Heißenbüttel}
\affiliation{
Institute of Solid State Theory, University of Münster, 48149 Münster, Germany}
\author{Thorsten Deilmann}
\affiliation{
Institute of Solid State Theory, University of Münster, 48149 Münster, Germany}
\author{Ermin Malic}

\affiliation{Department of Physics, Philipps-Universit{\"a}t Marburg, 35037 Marburg, Germany}
\affiliation{mar.quest|Marburg
Center for Quantum Materials and Sustainable Technologies,
35032 Marburg, Germany}

%%%%%%%%%%%%%%%%%%%%%%%%%%%%%%%%%%%%%%%%%%%%%%%%%%%%%%%%%%%%%%%%
\begin{abstract}
Two-dimensional magnetic semiconductors provide a unique platform for exploring the interplay between charge, lattice, spin, and light. The strong Coulomb interaction combined with long-range magnetic order enables efficient coupling between excitons and both lattice vibrations and spin excitations, thereby strongly influencing exciton dynamics and transport in these materials. Here, we develop a microscopic many-body theory of exciton--phonon and exciton--magnon interactions in mono- and bilayer CrSBr. We demonstrate that the quasi-one-dimensional character of excitons in CrSBr gives rise to strongly anisotropic exciton--phonon and exciton--magnon scattering, resulting in direction-dependent diffusion coefficients. Furthermore, we identify a distinct temperature dependence originating from one-phonon and two-magnon scattering processes, providing characteristic signatures of the underlying microscopic interactions. Our work establishes a microscopic framework for understanding exciton transport in magnetic van der Waals semiconductors and provides guidance for future optical spectroscopy and exciton transport experiments.
\end{abstract}
%%%%%%%%%%%%%%%%%%%%%%%%%%%%%%%%%%%%%%%%%%%%%%%%%%%%%%%%%%%%%%%%
\maketitle

%%%%%%%%%%%%%%%%%%%%%%%%%%%%%%%%%%%%%%%%%%%%%%%%%%%%%%%%%%%%%%%%
\section{Introduction}
%%%%%%%%%%%%%%%%%%%%%%%%%%%%%%%%%%%%%%%%%%%%%%%%%%%%%%%%%%%%%%%%
The discovery of intrinsic magnetism in atomically thin crystals has opened a new research field exploring the interplay between electronic, optical, and magnetic degrees of freedom~\cite{Jiang2021, adak2026excitons}. In contrast to conventional two-dimensional semiconductors, such as transition-metal dichalcogenides (TMDs)~\cite{mueller2018exciton,perea2022exciton, wang2018colloquium, lin2024moire}, magnetic van der Waals materials combine strong Coulomb interactions with long-range magnetic order, enabling combined optical, electronic and magnetic many-body phenomena. Among these materials, chromium sulfide bromide (CrSBr) has attracted particular attention due to its pronounced in-plane anisotropy, robust magnetic order facilitating strong exciton-magnon coupling ~\cite{Wilson2021, Heissenbuttel2025, Tschudin2024, Klein2023, TabatabaVakili2024}.

The optical response of CrSBr is dominated by tightly bound excitons whose wave functions inherit the highly anisotropic electronic band structure, giving rise to quasi-one-dimensional excitonic states~\cite{Heissenbuttel2025, Klein2023, Smiertka2026, Semina2025}. As a result, exciton dynamics is expected to be strongly direction dependent, making CrSBr a promising material platform for investigating microscopic mechanisms determining exciton relaxation and propagation.
\begin{figure}[b]
    \centering
    \includegraphics[width=\linewidth]{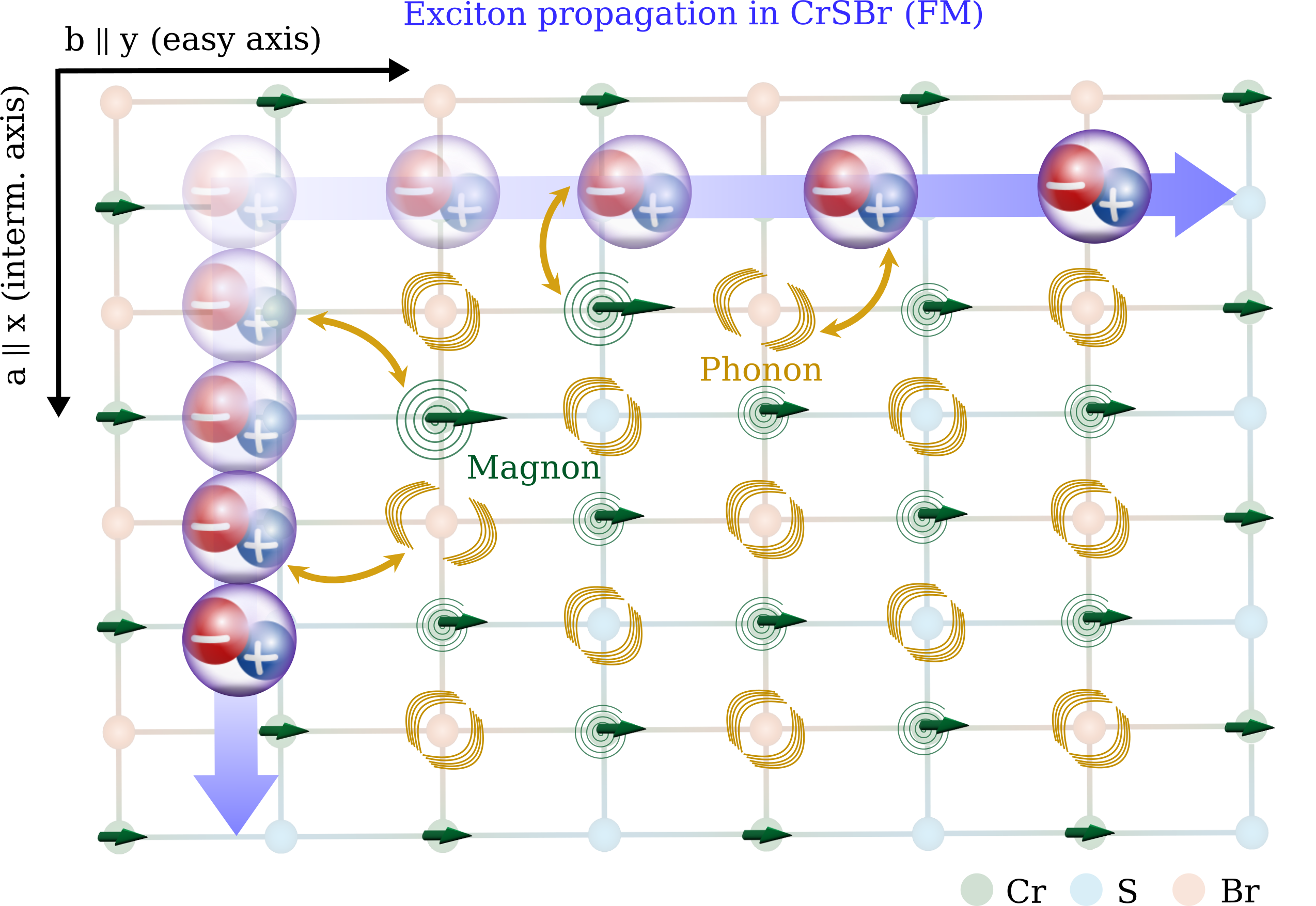}
    \caption{Schematic illustration of anisotropic exciton propagation in a ferromagnetic CrSBr monolayer in the presence of in-plane phonons and magnons. Magnons are localized on the Cr atoms (green) with the equilibrium magnetic moments being aligned along the b-axis (magnetic easy axis), and phonons are depicted as vibrating atoms (illustrated as outgoing soundwaves).}
    \label{fig1}
\end{figure}
Exciton transport is governed by scattering with elementary excitations. In TMD-based materials, exciton–phonon interactions determine exciton dephasing, linewidth broadening~\cite{moody2015intrinsic, selig2016excitonic, shree2018observation, brem2019intrinsic}, and diffusion~\cite{malic2023exciton, kulig2018exciton, zipfel2020exciton} at moderate excitation conditions. In CrSBr, excitons additionally couple to spin excitations, magnons, in its single-layer ferromagnetic (Fig.~\ref{fig1}) and multi-layer antiferromagnetic configurations, providing new scattering channels absent in non-magnetic systems~\cite{iakovlev2026boltzmann}. Recent experimental and theoretical studies have highlighted the important role of exciton-magnon interactions in CrSBr, including proposals that non-equilibrium magnon currents can drive efficient exciton transport~\cite{iakovlev2026boltzmann, Dirnberger2026}. These developments underscore the rich interplay between excitons and magnetic excitations, while also showing the importance of the still lacking microscopic understanding of how equilibrium exciton-phonon and exciton-magnon scattering determine the intrinsic anisotropic transport properties of excitons in the low-density regime.

In this work, we develop a microscopic theory of exciton--phonon and exciton--magnon interactions in mono- and bilayer CrSBr. Starting from a many-body Hamiltonian including excitons, phonons, and magnons, we derive the corresponding scattering rates and evaluate their impact on exciton spectral linewidths and diffusion. We demonstrate that the quasi-one-dimensional character of excitons leads to a strongly anisotropic scattering and transport behavior with a tenfold increase in the diffusion coefficient along the easy axis (b-axis) compared to the intermediate axis (a-axis) in CrSBr (Fig.~\ref{fig1}). Overall, our results provide a microscopic framework for understanding exciton transport in magnetic van der Waals semiconductors and establish a foundation for interpreting future optical and transport experiments in these emerging materials.\\

\section{Microscopic model}
In order to investigate exciton-phonon and exciton-magnon interactions in anisotropic magnetic semiconductors, we first define the many-body Hamiltonian. It contains kinetic contributions due to excitons, phonons and magnons ($H_0$) as well as the interaction between excitons and phonons ($H_{\mathrm{x-ph}}$) and excitons and magnons ($H_{\mathrm{x-m}}$). Focusing on the kinetic part of the exciton Hamiltonian and considering the energetically lowest 1\emph{s} exciton, we start with
\begin{equation}
\begin{split}
H_{0}=&\sum_{\mathbf{Q}}E_{\mathbf{Q}}X^{\dagger}_{\mathbf{Q}}X_{\mathbf{Q}}
+\sum_{i,\mathbf{q}}\hbar\Omega^{(\mathrm{ph})}_{i,\mathbf{q}}
b^{\dagger}_{i,\mathbf{q}}b_{i,\mathbf{q}}\\
&+\sum_{j,\mathbf{q}'}\hbar\Omega^{(\mathrm{m})}_{j,\mathbf{q}'}
m^{\dagger}_{j,\mathbf{q}'}m_{j,\mathbf{q}'} \, .
\label{h0}
\end{split}
\end{equation}
where the first term describes the kinetic motion of excitons with the
exciton dispersion  $E_{\mathbf{Q}}=\frac{\hbar^2 Q_x^2 }{2M^{(\mathrm{x})}_x}+\frac{\hbar^2 Q_y^2}{2M^{(\mathrm{x})}_y}+E_b$. Here, $\mathbf{Q}=(Q_x,Q_y)$ is the  center-of-mass (COM) momentum, $E_b$  the exciton binding energy, and $M^{(\mathrm{x})}_i=(m_{e,i}+m_{h,i})$  the direction-dependent exciton mass with $i=x,y$  and $m_{\lambda, i}$ as effective masses of electrons ($\lambda=e$) and holes ($\lambda=h$) considering a two-band model. The second and third terms in Eq. \ref{h0} contain the phonon and magnon dispersions, respectively. In the following, we discuss each term of the Hamiltonian in detail, starting with the electronic band structure that gives rise to the relevant excitonic transitions and the kinetic motion of excitons in CrSBr.   
\begin{figure}[t!]
\includegraphics[width=\columnwidth]{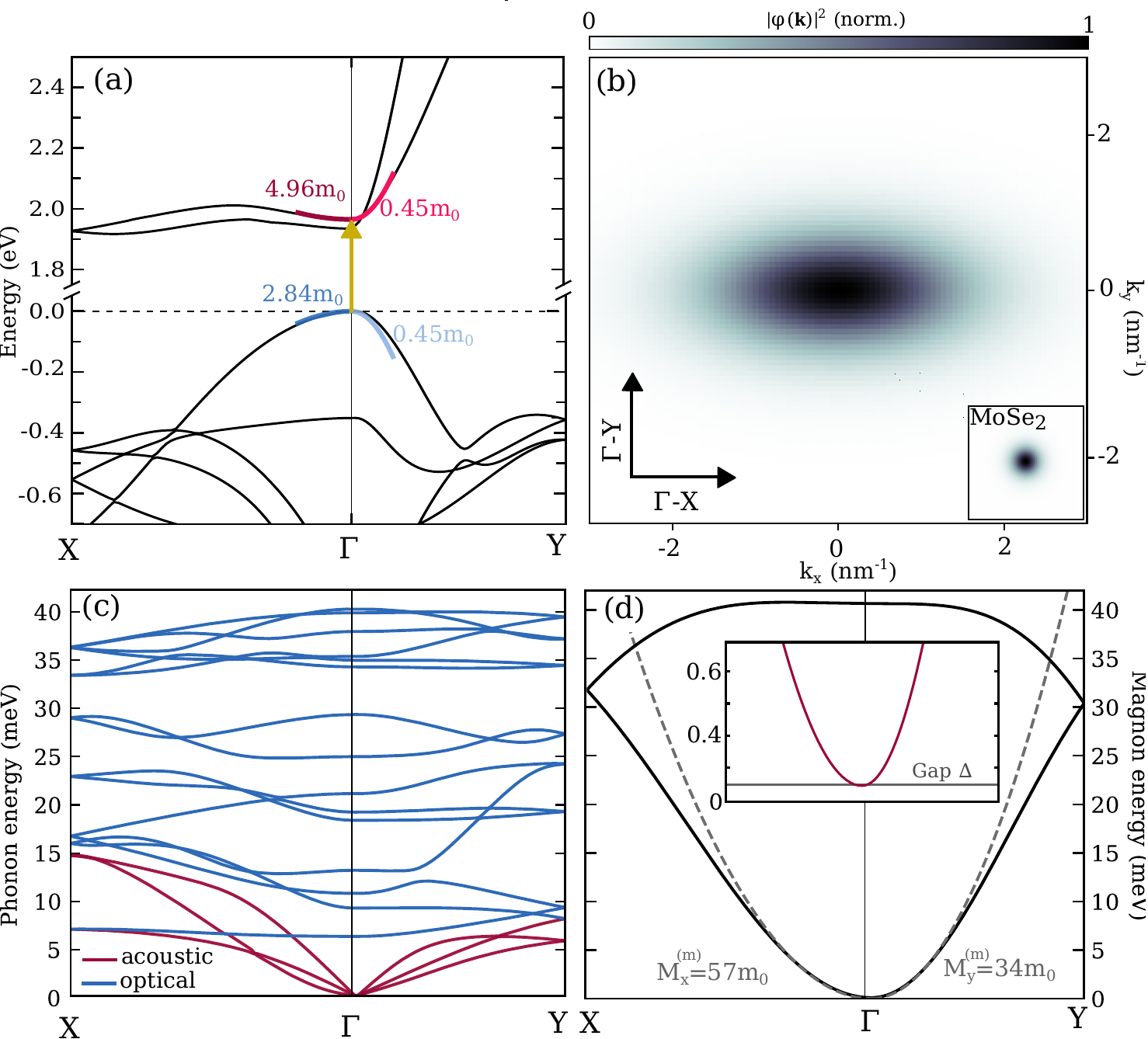}
\caption{Excitons, phonons and magnons in monolayer CrSBr. (a) Electronic band structure along the X-$\Gamma$-Y path in the rectangular Brillouin zone obtained from \emph{ab initio} calculations with the relevant carrier masses stated explicitly. (b) Anisotropic 1$s$ exciton wave function corresponding to the $(v,c+1)$ transition in (a) obtained by solving the Wannier equation -- in comparison with the isotropic 1\emph{s} wave function of  KK excitons in MoSe$_2$ monolayers (inset). The momentum scale is the same for both CrSBr and MoSe$_2$ wave functions. (c) Phonon dispersion with acoustic (red) and optical (blue) modes. (d) Ferromagnetic magnon dispersion with the low-energy magnon mode showing a parabolic momentum dependence with direction-dependent magnon masses $M^{(\mathrm{m})}_x\approx 57m_0$ and $M^{(\mathrm{m})}_y\approx{34} m_0$ (dashed grey lines) and a magnon gap $\Delta\approx 0.1$ meV around the $\Gamma$-point (inset). }
\label{fig2}
\end{figure}

%%%%%%%%%%%%%%%%%%%%%%%%%%%%%%%%%%%%%%%%%%%%%%%%%%%%%%%%%%%%%%%%
\subsection{Excitons}
%%%%%%%%%%%%%%%%%%%%%%%%%%%%%%%%%%%%%%%%%%%%%%%%%%%%%%%%%%%%%%%%
The electronic band structure of monolayer CrSBr is highly anisotropic in $\Gamma-X$ and $\Gamma-Y$ directions, see Fig.~\ref{fig2}(a). Due to its ferromagnetic nature, all low-energy bands of CrSBr carry the same spin polarization, thereby facilitating the formation of a variety of spin- and momentum-allowed exciton states~\cite{Smiertka2026, Qian2023}. We consider the exciton transition formed from the highest-lying valence band ($v$) and the second-lowest conduction band ($c+1$), marked in Fig.~\ref{fig2}(a). This transition is allowed within the dipole approximation assuming y-polarization ($y||b$) due to the irreducible representations associated with these bands (in contrast to the transition to the lowest-lying conduction band)~\cite{Klein2023, Semina2025, Heissenbuttel2025}. Such treatment is valid as hybridization effects between the $c$ and $c+1$ conduction bands are neglected in this work.  By solving the anisotropic Wannier equation for the bright $(v,c+1)$ transition, we obtain access to the exciton binding energy $E_b$ and the quasi-one-dimensional exciton wave function $\varphi$. The relevant direction-dependent reduced effective masses are obtained directly from our DFT+$GW$ calculations and carrier masses are provided explicitly in Fig.~\ref{fig2}(a). The dielectric screening is included as an anisotropic Rytova-Keldysh screening~\cite{keldysh1979coulomb, rytova1967screened, PhysRevB.84.085406} with screening radii extracted from the DFT-calculated dielectric function~\cite{Klein2023} in the long wavelength limit (see Supplemental Section I for details). In Fig.~\ref{fig2}(b), we show the corresponding momentum-dependent exciton wave function of the bright $\Gamma\Gamma$ exciton in a hBN-encapsulated CrSBr monolayer. We obtain a binding energy $E_b\approx 227$ meV and a highly anisotropic wave function of the 1$s$  exciton that extends along the $\Gamma-X$ direction ($a$-axis). The anisotropy predominantly stems from the anisotropic reduced exciton mass ($\mu_x/\mu_y\approx 8$) with only a minor contribution from the direction-dependent screening (radii $r_x/r_y\approx{2}$) of the material. For comparison, we also show the 1$s$ exciton wave function in monolayer MoSe$_2$, see the inset in Fig.~\ref{fig2}(b). In contrast to MoSe$_2$, the exciton wave function in CrSBr is not rotationally symmetric, but exhibits a direction-dependent Bohr radius $a_{B,i}$  ($i=x,y$) such that $a_{B,y}/a_{B,x}\approx 2$. As expected from the Wannier model, the exciton wave function is well-contained within the first Brillouin zone ($|\Gamma-X|\approx 9$ $\mathrm{nm}^{-1}$ and $|\Gamma-Y|\approx 7$ $\mathrm{nm}^{-1}$). We note that complementary many-body approaches based on the Bethe-Salpeter equation have shown that low-energy excitons in CrSBr exhibit a mixed Wannier-Frenkel character \cite{Klein2023, Smiertka2026}. In Fig. S1 in Supplemental Section II, we therefore provide the lowest-lying exciton wave function as obtained from solving the Bethe-Salpeter equation (BSE). We find that the BSE-calculated exciton wave function is limited to the first Brillouin zone, but extends a bit further in momentum space compared to the wave function obtained from solving the anisotropic Wannier equation (Fig. 2b)). The difference in localization is attributed to the inclusion of several flatter bands in the BSE calculation. Since the detailed structure of the exciton wave function is not in the focus of this work, we employ the Wannier equation approach to investigate exciton-phonon and exciton-magnon interactions in CrSBr.

%%%%%%%%%%%%%%%%%%%%%%%%%%%%%%%%%%%%%%%%%%%%%%%%%%%%%%%%%%%%%%%%
\subsection{Phonons and magnons}
%%%%%%%%%%%%%%%%%%%%%%%%%%%%%%%%%%%%%%%%%%%%%%%%%%%%%%%%%%%%%%%%
Having discussed excitons in CrSBr, we proceed to the phonon and magnon dispersion, the latter arising from the intrinsic ferromagnetic order of the material. The phonon Hamiltonian is given by the second term in Eq.~\eqref{h0}, where $b^{(\dagger)}_{i, \mathbf{q}}$ are bosonic phonon operators with  $i$ denoting the phonon mode and $\mathbf q$ the phonon momentum. The phonon dispersion $\hbar\Omega^{(\mathrm{ph})}_{i, \mathbf{q}}$ is obtained from \emph{ab initio} calculations and we find that CrSBr monolayers exhibit 18 different phonon modes, 3 acoustic (ZA, LA and TA modes corresponding to out-of-plane, in-plane longitudinal and transversal vibrations, respectively) and 15 optical modes (Fig.~\ref{fig2}(c)). The phonon dispersion is less anisotropic than the electronic band structure around the $\Gamma$-point with speeds of sound ranging from $v_{\Gamma-X}=1860$ m/s ($v_{\Gamma-X}=4960$ m/s) to  $v_{\Gamma-Y}=2540$ m/s ($v_{\Gamma-Y}=5040$ m/s) for transversal (longitudinal) acoustic modes. Additional details on the \emph{ab initio} calculations are provided in Supplemental Section III. 

Besides quantized lattice vibrations, CrSBr being a ferromagnetic (antiferromagnetic) material in its monolayer (bilayer and bulk) form, hosts spin excitations, i.e. magnons. We obtain the magnon dispersion of CrSBr monolayers by diagonalizing a ferromagnetic magnon Hamiltonian that includes nearest-neighbor exchange terms as well as lattice anisotropy terms (see Supplemental Section IV). Inclusion of next-nearest neighbors and dipolar interactions between spins are assumed to provide small quantitative corrections to the magnon dispersion \cite{iakovlev2026boltzmann}. This results in the third term in Eq.~\eqref{h0} introducing the bosonic magnon operators $m^{(\dagger)}_{j, \mathbf{q}'}$, with $j$ denoting the magnon modes and $\mathbf q$ the magnon momentum. As shown in Fig. 2 (d), two magnon modes are obtained due to the two magnetic Cr atoms in the CrSBr unit cell, and the magnon energies lie in a similar range as those of phonons. The minimum of the low-energy magnon mode resides at $\hbar \Omega^{(m)}_{\mathbf{q}=\Gamma}\approx{0.1 }$ meV and the magnon dispersion is gapped due to the weak lattice anisotropy (inset in Fig.~\ref{fig2}(d)). In addition, the low-energy magnon mode exhibits a parabolic dispersion around the $\Gamma$-point, such that magnon masses $M^{(\mathrm{m})}_{x}\approx{57m_0}$ and $M^{(\mathrm{m})}_{y}\approx{34m_0}$ can be extracted (see dashed gray lines in Fig.~\ref{fig2}(d)). The parabolic approximation of the magnon dispersion (including the magnon gap) is applied in the numerical calculations of the exciton-magnon scattering rates. The qualitatively different behaviour of low-energy phonons and magnons as a function of momentum is of crucial importance for their scattering with excitons, as discussed below.  

%%%%%%%%%%%%%%%%%%%%%%%%%%%%%%%%%%%%%%%%%%%%%%%%%%%%%%%%%%%%%%%%
\subsection{Exciton scattering theory}
%%%%%%%%%%%%%%%%%%%%%%%%%%%%%%%%%%%%%%%%%%%%%%%%%%%%%%%%%%%%%%%%
The interaction between excitons and phonons contributes to the non-radiative recombination of excitons in atomically thin semiconductors and  gives rise to a homogeneous broadening of exciton resonances with temperature~\cite{moody2015intrinsic, brem2019intrinsic}. In magnetic semiconductors, such as CrSBr, it is expected that also the magnetic order influences the exciton linewidth via exciton-magnon scattering~\cite{iakovlev2026boltzmann}. 
The exciton-phonon Hamiltonian reads
\begin{equation}
H_{\mathrm{x-ph}}=\sum_{j, \mathbf{q}}G^{(\mathrm{ph})}_{j, \mathbf{q}}(b_{j, \mathbf{q}}+b^{\dagger}_{j, -\mathbf{q}})X^{\dagger}_{\mathbf{Q+q}}X_{\mathbf{Q}} \ , 
\end{equation}
introducing the exciton-phonon coupling $G^{(\mathrm{ph})}_{j, \mathbf{q}}$ that depends on carrier-phonon matrix elements and exciton wave function overlaps. The former are treated within the deformation potential approximation and estimated by a single (averaged) deformation potential for optical and in-plane longitudinal and transversal acoustic phonons, respectively. The relevant deformation potential parameters needed for evaluating the phonon-induced dephasing are obtained as fitting parameters to recent experimental linewidth data on CrSBr~\cite{Dirnberger2026} (see Supplemental Section V for details).   Given the exciton-phonon Hamiltonian, the corresponding phonon-induced dephasing is derived using a Heisenberg equation-of-motion-approach and the exciton semiconductor Bloch equation for the microscopic polarization within a second-order Born-Markov approximation. The dephasing reads~\cite{selig2016excitonic}
\begin{equation}
\gamma_{\mathrm{ph}}(\mathbf{Q})=\pi\sum_{j, \mathbf{q}, \pm}|G^{(\mathrm{ph})}_{j, \mathbf{q}}|^2 (n^{(\mathrm{ph})}_{j, \mathbf{q}}+\frac{1}{2}\pm \frac{1}{2})\delta(\epsilon^{\pm}_{j, \mathbf{Q},\mathbf{q}}) \ , 
\label{gammaph}
\end{equation}
where $\epsilon^{\pm}_{j, \mathbf{Q},\mathbf{q}}=E_{\mathbf{Q+q}}-E_{\mathbf{Q}}\pm \hbar\Omega^{(\mathrm{ph})}_{j, \mathbf{q}}$ and $n^{(\mathrm{ph})}_{j, \mathbf{q}}$ is taken as a thermalized Bose-Einstein distribution. Furthermore, one-phonon emission ($+$) as well as phonon absorption ($-$) contribute to the dephasing. In this work, the energy-conserving $\delta$-function entering  exciton-phonon and exciton-magnon scattering rates is approximated by a Gaussian with a fixed width $\Gamma = 1$~meV. Besides improving numerical convergence, this phenomenological broadening effectively accounts for higher-order scattering processes beyond the lowest-order treatment by relaxing the strict energy-conservation condition~\cite{rossi2002theory}. For  $Q=0$, the phonon-induced dephasing reduces to half the full width at half maximum of exciton broadening.

In order to describe the coupling between excitons and magnons, we start from a Hamiltonian expressed in terms of spin operators due to itinerant electrons and localized magnetic moments~\cite{PhysRevB.1.4474, PhysRevB.97.115401}. By expressing the Hamiltonian in terms of electronic operators and transforming the spin operators of the localized magnetic moments to magnon operators by means of a Holstein-Primakoff transformation~\cite{holstein1940field}, an electronic Hamiltonian is obtained (see Supplemental Section VI for a detailed derivation). In the excitonic picture, the resulting exciton-magnon Hamiltonian reads
\begin{equation}
H_{\mathrm{x-m}}=\sum_{j, \mathbf{Q}, \mathbf{q}, \mathbf{q}'}G^{(\mathrm{m})}_{\mathbf{q}'} m^{\dagger}_{j, \mathbf{q}-\mathbf{q}'}m_{j, \mathbf{q}}X^{\dagger}_{\mathbf{Q}+\mathbf{q}'}X_{\mathbf{Q}} \ , 
\label{2magham}
\end{equation}
where the exciton-magnon coupling is given by $G^{(m)}_{\mathbf{q}}=\frac{\mathcal{A}_0}{2}[I_e F(\boldsymbol{\beta}\cdot\mathbf{q})-I_h F(-\boldsymbol{\alpha}\cdot\mathbf{q})]$ with the carrier-magnon couplings $I_{\lambda}$, $\lambda=e,h$, the exciton form factor $F(\mathbf{q})=\sum_{\mathbf{k}}\varphi^*_{\mathbf{k+q}}\varphi_{\mathbf{k}}$, the carrier mass ratio $\boldsymbol{\beta}=1-\boldsymbol{\alpha}=(\frac{m_{e,x}}{m_{e,x}+m_{h,x}}, \frac{m_{e,y}}{m_{e,y}+m_{h,y}})$ and the magnetic unit cell area $\mathcal{A}_0$. The form of the exciton-magnon coupling is qualitatively similar to that of the exciton-phonon coupling as both matrix elements depend on electronic matrix elements weighted by exciton form factors (see Supplemental Section V). 

Here, we take the carrier-magnon couplings $I_{\lambda}$ as mode-independent. To the best of our knowledge, microscopic calculations of the carrier--magnon couplings for CrSBr have not been reported. We therefore adopt an estimate for the hole--magnon coupling, $I_h\approx 50$ meV, based on the previously obtained energy shift due to the hole-magnon coupling in the related ferromagnet EuCd$_2$As$_2$~\cite{Jo2021}. Since the exciton--magnon coupling strength in CrSBr is presently not known from microscopic calculations, we refrain from making quantitative comparisons between the contributions of exciton--phonon and exciton--magnon scattering to exciton linewidths and transport. Instead, we focus on the robust qualitative trends and characteristic signatures associated with each scattering mechanism.

The exciton-magnon Hamiltonian is of second order in magnon operators $m^{(\dagger)}$. In fact, since creating or annihilating a single magnon corresponds to a spin-flip process, a one-magnon scattering process is only possible if the interaction involves scattering between spin-bright and spin-dark excitons or if an external magnetic field is applied. Since exciton-magnon interaction for bright excitons is vanishing at first order and is only finite at second order in the magnon operators, we hence expect the overall interaction to be enhanced with an external magnetic field, in agreement with recent experimental studies~\cite{adak2026microwave, Datta2025}. In particular, in bilayer and bulk structures, exciton-magnon interaction gives rise to a significant magnetic-field dependent energy renormalization of the exciton energy and a splitting of exciton states at antiferromagnetic and ferromagnetic configurations~\cite{Wilson2021}. Here, we focus on the corresponding magnon-induced dephasing derived from the two-magnon Hamiltonian. Using the Heisenberg equation-of-motion approach (Supplemental Section V), we obtain the magnon-induced dephasing: 
\begin{equation}
\gamma_{\mathrm{m}}(\mathbf{Q})=\pi\sum_{j, \mathbf{q}, \mathbf{q}'}|G^{(\mathrm{m})}_{\mathbf{q}-\mathbf{q}'}|^2 n^{(\mathrm{m})}_{j, \mathbf{q}}(1+n^{(\mathrm{m})}_{j, \mathbf{q}'})\delta(\epsilon_{j, \mathbf{Q}, \mathbf{q}, \mathbf{q}'}) \ , 
\label{2mag}
\end{equation}
where $\epsilon_{j, \mathbf{Q}, \mathbf{q}, \mathbf{q}'}=E_{\mathbf{Q+q-q'}}-E_{\mathbf{Q}}+\hbar\Omega^{(\mathrm{m})}_{j, \mathbf{q}'}-\hbar\Omega^{(\mathrm{m})}_{j, \mathbf{q}}$ and $n^{(\mathrm{m})}_{j, \mathbf{q}}$ is the thermalized Bose-Einstein distribution of magnons.

%%%%%%%%%%%%%%%%%%%%%%%%%%%%%%%%%%%%%%%%%%%%%%%%%%%%%%%%%%%%%%%%
\section{Phonon- and magnon-induced dephasing}
\begin{figure}[t]
\includegraphics[width=\linewidth]{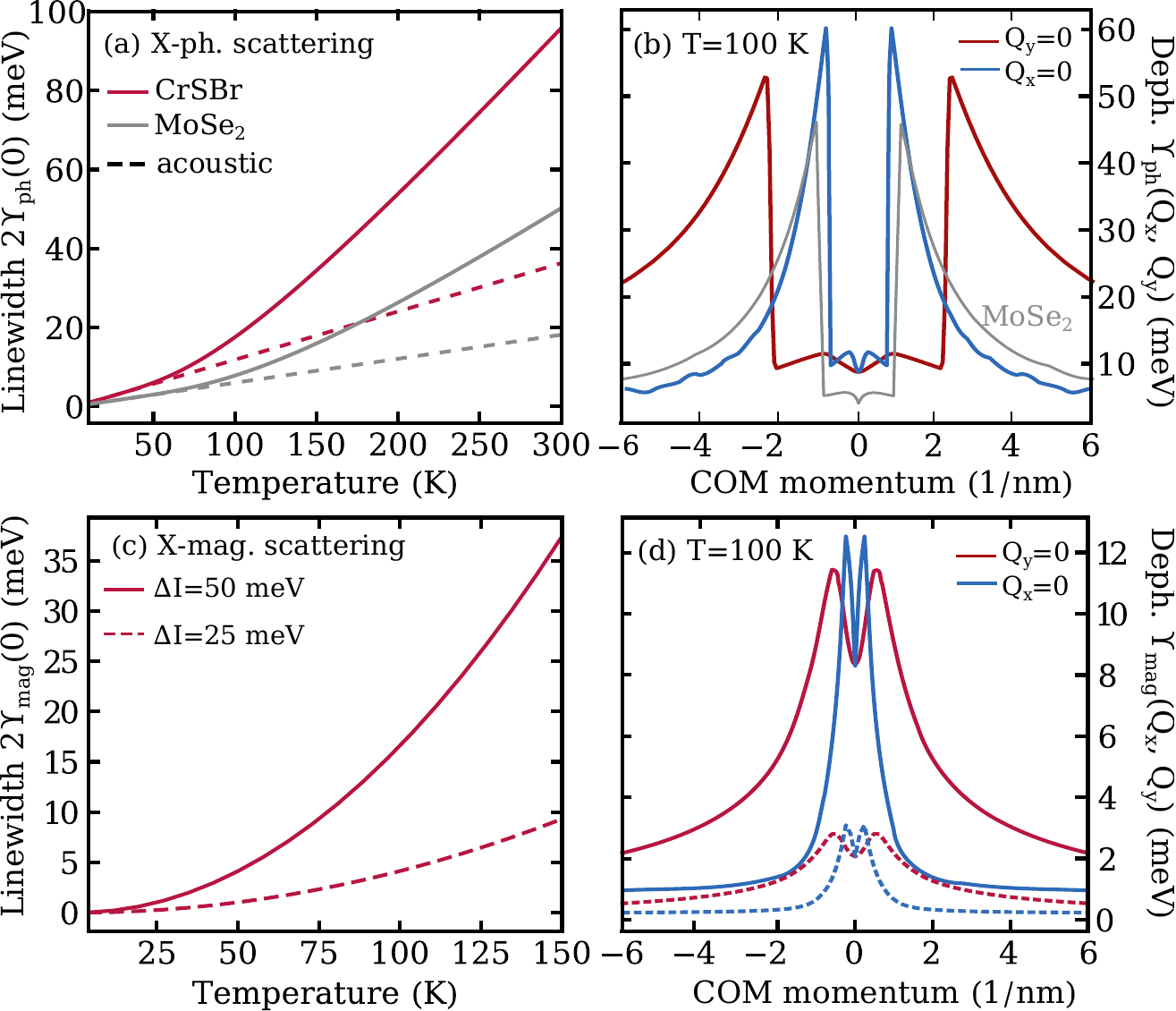}
\caption{Phonon- and magnon-induced dephasing in monolayer CrSBr. (a) Non-radiative exciton broadening from exciton–phonon scattering, showing contributions from acoustic phonons only (dashed) and from both acoustic and optical phonons (solid). For comparison, the linewidth of  MoSe$_2$ monolayers is also shown (gray). (b) Direction-dependent momentum-resolved phonon-induced dephasing in CrSBr (red line: scattering along $Q_x$, blue line: scattering along $Q_y$) and isotropic scattering in monolayer MoSe$_2$. (c) Non-radiative exciton broadening due to exciton-magnon scattering for different carrier-magnon couplings, $\Delta I=|I_e-I_h|$. (d) Direction-dependent momentum-resolved magnon-induced dephasing. The solid lines correspond to $\Delta I=50$~meV and vanishing electron-magnon coupling, whereas the dashed lines corresponds to the case $\Delta I=25$ meV.}
\label{fig3}
\end{figure}

%%%%%%%%%%%%%%%%%%%%%%%%%%%%%%%%%%%%%%%%%%%%%%%%%%%%%%%%%%%%%%%%
In the following, we evaluate the phonon- and magnon-induced dephasings in hBN-encapsulated CrSBr monolayers. In Fig.~\ref{fig3}(a), we show the temperature-dependent exciton linewidth (non-radiative broadening) due to exciton-phonon scattering evaluated by solving Eq.~\eqref{gammaph}. At low temperatures, scattering with acoustic phonons is most efficient, resulting in a linear dependence of the linewidth with respect to temperature (dashed lines), whereas at elevated temperatures, emission of optical phonons becomes possible, leading to a non-linear increase of the exciton linewidth as commonly found for two-dimensional semiconductors (solid lines)~\cite{selig2016excitonic, shree2018observation, brem2019intrinsic}. We show, however, that the linewidth of CrSBr is generally much larger than the linewidths in isotropic TMD materials, in particular considering here the representative MoSe$_2$ monolayers (gray lines). Since the extracted deformation potentials and phonon energies are quite similar for these material systems (Supplemental Section V), we ascribe the larger linewidth to the greater geometric exciton mass of CrSBr with $\bar{M}^{(\mathrm{x})}=\sqrt{M^{(\mathrm{x})}_x M^{(\mathrm{x})}_y}\approx{2.6}m_0$ compared to $\bar{M}^{(\mathrm{x})}\approx 1.1m_0$ for MoSe$_2$ \cite{kormanyos2015k}. A larger exciton mass enhances the phase-space for efficient exciton-phonon scattering with acoustic phonons reflected by $\gamma_{\mathrm{ph}}(T)|_{\mathrm{ac}}\propto \bar{M}^{(\mathrm{x})}T$~\cite{shree2018observation} with $T$ being the temperature. 

In Fig.~\ref{fig3}(b), the fully momentum-dependent phonon-induced dephasing is shown at the fixed temperature $T=100$ K, revealing anisotropic exciton-phonon scattering in CrSBr. In particular, the drastic increase of the dephasing at finite COM momenta corresponds to momenta at which optical phonon emission becomes efficient in the considered directions, i.e. $Q_{i, \mathrm{op}}=\sqrt{2M^{(\mathrm{x})}_i\Omega_{\mathrm{op}}/\hbar}$ with $i=x,y$ and $\hbar\Omega_{\mathrm{op}}\approx{25}$ meV being the averaged optical phonon energy in CrSBr obtained within the Einstein approximation (Fig.~\ref{fig2}(c)). Since in CrSBr we find that $M^{(\mathrm{x})}_x/M^{(\mathrm{x})}_y\approx 8.7$ for the considered exciton at the $\Gamma$-point, optical phonon emission becomes efficient (and at a smaller COM momentum) along the $Q_y$-direction (blue lines) than along the $Q_x$-direction (red lines). In comparison, exciton-phonon scattering in MoSe$_2$ is fully isotropic (gray lines). 

In Fig.~\ref{fig3}(c), we find a parabolic temperature dependence of the exciton linewidth due to exciton-magnon scattering, which  deviates from the linear temperature dependence obtained from acoustic exciton-phonon scattering. This is due to the two-magnon nature of the scattering process: since the exciton-magnon scattering integral (Eq.~\eqref{2mag}) is proportional to $n^{(\mathrm{m})}_{\mathbf{q}}n^{(\mathrm{m})}_{\mathbf{q}'}\propto T^2$ at low temperatures, the non-linear temperature dependence of the linewidth is expected. In fact, it can be shown more generally that $\gamma_{\mathrm{m}}(0)\propto \frac{(k_B T)^2}{\Delta}$ assuming that the magnon masses are much larger than the exciton masses and gap $\Delta \ll k_B T$, see Supplemental Section VI. This result also highlights the role of the magnon gap as an infrared cutoff. In the gapless limit, the long-wavelength thermal occupation of magnons with a quadratic dispersion (inset in Fig.~\ref{fig2}(d)) leads to an infrared divergence of the magnon-induced dephasing, analogous to the infrared divergences underlying the Mermin–Wagner theorem in two-dimensional magnets. A finite gap regularizes these long-wavelength fluctuations yielding a finite dephasing~\cite{PhysRevB.93.155413}. 

Here, we have fixed the hole-magnon coupling $I_h\approx 50$ meV and consider two different cases for the electron-magnon coupling $I_e$ entering the exciton-magnon interaction in Eq.~\eqref{2magham}. The value for $I_h$ is estimated based on the previously obtained energy shift due to the hole-magnon coupling in the ferromagnet EuCd$_2$As$_2$~\cite{Jo2021}. Depending on the strength of the carrier-magnon couplings, the magnon-induced dephasing can become comparable to the phonon-induced dephasing. Since the exciton-magnon interaction depends on the difference between electron-magnon and hole-magnon couplings (Eq.~\eqref{2magham}), the magnon-induced exciton linewidth is substantially reduced if $I_e\approx I_h$. Note, however, that the qualitative quadratic temperature dependence remains unchanged, irrespective of the ratio $I_e/I_h$. This can be understood from the behavior of the exciton form factors entering the exciton–magnon interaction: they decay on the scale of the inverse exciton Bohr radius, which is much slower than the momentum-dependent variation of the magnon population in the considered temperature regime. Consequently, the exciton–magnon coupling can be approximated as a momentum-independent constant. A modification of the carrier–magnon coupling therefore primarily results in a rescaling of the exciton linewidth, or more generally, of the magnon-induced dephasing (cf. the solid and dashed lines in Figs.~\ref{fig3}(c)-(d)).

We find that the magnon-induced dephasing in CrSBr is anisotropic and exhibits a non-monotonous dependence as a function of momentum, see Fig.~\ref{fig3}(d). Generally, the phase-space for exciton-magnon scattering is enhanced with increased COM momentum resulting in a larger dephasing for small COM momenta. In order for energy to be conserved in the two-magnon scattering process at larger COM momenta, high-energy magnons with larger momenta participate in the scattering process and these are weakly occupied in the considered temperature regime resulting in an overall decrease of the dephasing (cf. Eq.~\eqref{2mag}). In the anisotropic case of CrSBr, where $M^{(\mathrm{x})}_x/M^{(\mathrm{x})}_y\approx 8$, the exciton energy $E_{Q_x}$ grows slower with momentum compared to $E_{Q_y}$. Hence, low-energy magnons contribute to an efficient scattering process for a broader range of COM momenta when scattering with excitons along the $Q_x$-direction (red lines in Fig.~\ref{fig3}(d)) compared to the scattering along the $Q_y$-direction (blue lines in Fig.~\ref{fig3}(d)). Furthermore, because the magnon dispersion is quadratic in momentum (dashed lines in Fig.~\ref{fig2}(d)), whereas the in-plane acoustic phonon dispersion is linear, the magnon population decays more rapidly with increasing momentum, scaling as $n^{(\mathrm{m})}_{\mathbf{q}}\sim \mathrm{e}^{-cq^2}$ compared to $n^{(\mathrm{ph})}_{\mathbf{q}}\sim \mathrm{e}^{-c'q}$ for acoustic phonons. Since the magnon and phonon populations directly enter the corresponding dephasings (Eq. \eqref{gammaph} and \eqref{2mag}) and predominantly determine the momentum dependence of the scattering integrands, the magnon-induced dephasing exhibits a more rapid momentum-dependent decay than the phonon-induced dephasing.
Note that exciton-phonon and exciton-magnon scattering both contribute to the non-radiative broadening of exciton resonances below the Curie temperature, $T_C\approx {146}$ K, under which the material is ferromagnetic. If exciton-magnon coupling is substantially larger than exciton-phonon coupling, it is expected that the non-radiative broadening and the overall dephasing could  be reduced above the Curie temperature and a non-monotonous temperature dependence could be observed around the magnetic phase transition.    

%%%%%%%%%%%%%%%%%%%%%%%%%%%%%%%%%%%%%%%%%%%%%%%%%%%%%%%%%%%%%%%%
\section{Anisotropic exciton propagation}
%%%%%%%%%%%%%%%%%%%%%%%%%%%%%%%%%%%%%%%%%%%%%%%%%%%%%%%%%%%%%%%%
The anisotropic exciton-phonon and exciton-magnon scattering discussed above provides the microscopic basis for the modelling of exciton transport in CrSBr in the linear regime of small exciton, phonon and magnon densities. Specifically, the associated dephasings determine the relaxation of propagating excitons and enter directly into the diffusion coefficient. We therefore proceed by investigating how these scattering mechanisms shape the anisotropic propagation of excitons in CrSBr. The $i$-th component of the diffusion coefficient in a two-dimensional anisotropic semiconductor reads~\cite{Dirnberger2026, chang2026ultrafast, thompson2022anisotropic, hess1996maxwell}
\begin{equation}
D_i=\sum_{\mathbf{Q}}\tau_{\mathbf{Q}}v^2_{i, \mathbf{Q}} \frac{N_{\mathbf{Q}}}{\mathcal{Z}} 
\label{diffusion}
\end{equation}
where $v_{i, \mathbf{Q}}=\hbar Q_i/M^{(\mathrm{x})}_{i}$ is the group velocity associated with an exciton with COM momentum $\mathbf{Q}$ along the $i$-th direction and $\tau_{\mathbf{Q}}=\hbar/(2\gamma_{\mathbf{Q}})$ is the scattering time. Furthermore, $N_{\mathbf{Q}}$ is the exciton occupation with the partition function $\mathcal{Z}=\sum_{\mathbf{Q}}N_{\mathbf{Q}}$. The exciton occupation is estimated by a thermalized Boltzmann occupation. In this thermalized limit and assuming state-independent scattering times, the diffusion coefficient reduces to the semiclassical expression $D_i=k_B T\tau/M_i$ \cite{PhysRevLett.120.207401}. The dephasing $\gamma_{\mathbf{Q}}$ determining the scattering time formally includes contributions from both exciton-phonon and exciton-magnon scattering processes. However, we treat these contributions separately, which allows us to identify the characteristic qualitative trends in exciton diffusion associated with the individual scattering mechanisms.

%%%%%%%%%%%%%%%%%%%%%%%%%%%%%%%%%%%%%%%%%%%%%%%%%%%%%%%%%%%%%%%%
\subsection{Monolayer CrSBr}
\begin{figure}[t]
    \centering
    \includegraphics[width=\columnwidth]{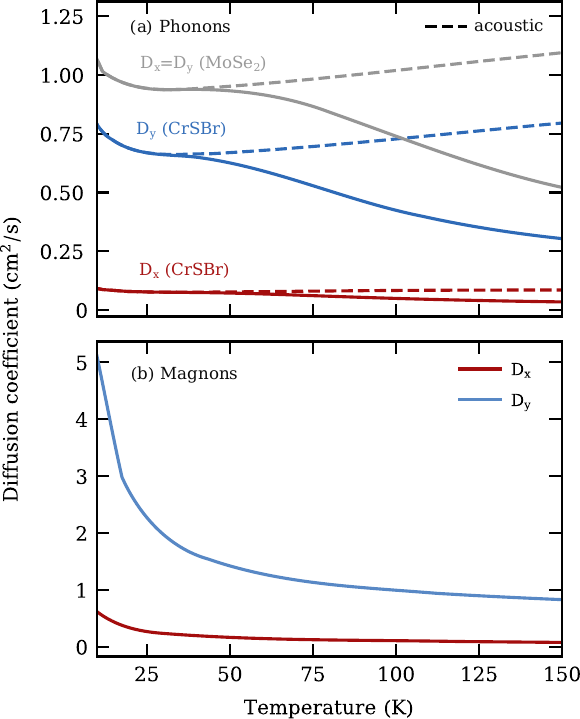}
    \caption{Anisotropic exciton propagation in CrSBr monolayers. (a) Direction-dependent diffusion coefficient due to exciton-phonon scattering in CrSBr (red and blue) in comparison with the isotropic diffusion coefficient of monolayer MoSe$_2$ (gray). The contributions from acoustic phonons are marked as dashed lines, whereas the solid lines include contributions from both acoustic and optical phonons. (b) Exciton diffusion due to two-magnon-exciton scattering (carrier-magnon couplings set such that $\Delta I=|I_e-I_h|=50$ meV).}
    \label{fig4}
\end{figure}

%%%%%%%%%%%%%%%%%%%%%%%%%%%%%%%%%%%%%%%%%%%%%%%%%%%%%%%%%%%%%%%%
Having microscopic access to the phonon- and magnon-induced dephasing via Eq.~\eqref{gammaph} and Eq.~\eqref{2mag}, we evaluate the temperature-dependent exciton diffusion coefficient (Eq.~\eqref{diffusion}) for monolayer CrSBr separating contributions from exciton-phonon (Fig.~\ref{fig4}(a)) and exciton-magnon scattering (Fig.~\ref{fig4}(b)). We find that the temperature dependence of the diffusion coefficients due to exciton-phonon scattering in monolayer CrSBr (blue and red) and MoSe$_2$ (gray) are qualitatively similar. The larger diffusion coefficient in monolayer MoSe$_2$ is due to the slightly weaker exciton-phonon scattering and smaller exciton mass $M^{(\mathrm{x})}_{x/y}=1.1m_0$ in MoSe$_2$. We find that acoustic phonons (dashed lines) contribute to a weakly temperature-dependent diffusion coefficient. As temperature is increased, exciton states with higher momentum and thereby larger group velocities are occupied enhancing exciton propagation. At the same time, exciton-phonon scattering is increased with temperature (Fig.~\ref{fig3}(a)), counteracting the increase in group velocities and resulting in a nearly temperature-independent diffusion coefficient. This can also be shown analytically by taking the scattering time as constant, i.e., $\tau_{\mathrm{ph}}\propto \gamma^{-1}_{\mathrm{ph}}(\mathbf{0})\propto 1/T$ for acoustic phonons \cite{shree2018observation}. In addition, since $\langle v^2_{i, \mathbf{Q}}\rangle =\sum_{\mathbf{Q}} v^2_{i, \mathbf{Q}} N_{\mathbf{Q}}/\mathcal{Z}=k_B T/M^{(\mathrm{x})}_i $ as expected from the equipartition theorem, it directly follows that the diffusion coefficient $D_i$ (Eq. \eqref{diffusion}) is approximately independent of temperature. 
The inclusion of optical phonons (solid lines) results in a decay of the exciton propagation above 50 K when higher-momentum exciton-phonon scattering and optical phonon emission become efficient (Fig.~\ref{fig3} (b)). This follows from the stronger temperature dependence of exciton-phonon scattering with optical phonons compared to acoustic phonons (Fig.~\ref{fig3}(a)), i.e., whereas $\gamma_{\mathrm{ph}}|_{\mathrm{ac}}\propto T$ for acoustic phonons, it instead holds that $\gamma_{\mathrm{ph}}|_{\mathrm{op}}\propto (\mathrm{exp}(\hbar\Omega_{\mathrm{op}}/(k_B T))-1)^{-1}$ for optical phonons.  Hence, the increase in diffusion due to the population of exciton states with large group velocities is overcompensated by the stronger scattering with optical phonons.

Furthermore, in contrast to monolayer MoSe$_2$ (gray lines), the diffusion coefficient in CrSBr is highly anisotropic with $D_x\approx 0.09$ $\mathrm{cm}^2/$s ($D_x\approx 0.05$ $\mathrm{cm}^2/$s) and $D_y\approx 0.75$ $\mathrm{cm}^2/$s ($D_y\approx 0.4$ $\mathrm{cm}^2/$s) at T=10 K (T=100 K) corresponding to propagation along the $a$- and $b$-axes, respectively (Fig.~\ref{fig1}). The anisotropy in the exciton transport partially comes from the anisotropic exciton-phonon scattering rates, but predominantly follows from the substantially different group velocities in $x$- and $y$-directions. In particular, it then holds that the ratio of exciton diffusion coefficients is approximately temperature-independent with $\frac{D_y}{D_x}\approx \frac{M^{(\mathrm{x})}_x}{M^{(\mathrm{x})}_y}\approx 8.7$ (disregarding the small correction due to the anisotropy in the exciton-phonon scattering rate, cf. Fig~\ref{fig3}(b)).

The temperature dependence of the diffusion coefficient due to exciton-magnon scattering (Fig.~\ref{fig4}(b)) is substantially stronger than in the case of exciton-phonon scattering (Fig.~\ref{fig4}(a)). This follows from the two-magnon nature of the exciton-magnon scattering: since the magnon-induced dephasing scales quadratically with temperature (Fig.~\ref{fig3}(c)) and thereby the scattering time $\tau_{\mathrm{m}}\propto 1/T^2$, it holds that the diffusion coefficient $D_{i}=k_B T\tau_{\mathrm{m}}/M^{(\mathrm{x})}_i\propto \frac{1}{T}$ assuming a state-independent scattering time and the thermalized regime \cite{wagner2021nonclassical, erkensten2026impact}. Hence, the exciton diffusion coefficient due to exciton-magnon scattering is approximately inversely proportional to temperature and is monotonously decaying with temperature (Fig. ~\ref{fig4}(b)). Within the coupling strengths employed here, exciton-magnon scattering yields larger diffusion coefficients than exciton-phonon scattering because the corresponding scattering rates are smaller (see Fig.~\ref{fig3}(a)-(b) and Fig.~\ref{fig3}(c)-(d)). Similar to the case of exciton-phonon scattering, the anisotropy in the magnon-induced dephasing (Fig.~\ref{fig3}(d)) combined with the large anisotropy in the group velocities, results in a strongly direction-dependent diffusion coefficient due to exciton-magnon scattering. Overall, we find that both exciton-magnon and exciton-phonon scattering result in a temperature-dependent and highly anisotropic exciton propagation in CrSBr.    

%%%%%%%%%%%%%%%%%%%%%%%%%%%%%%%%%%%%%%%%%%%%%%%%%%%%%%%%%%%%%%%%
\subsection{Bilayer CrSBr}
%%%%%%%%%%%%%%%%%%%%%%%%%%%%%%%%%%%%%%%%%%%%%%%%%%%%%%%%%%%%%%%%
Having discussed exciton-phonon and exciton-magnon coupling in the ferromagnetic CrSBr monolayer configuration, we proceed to investigate the antiferromagnetic CrSBr bilayer. We do not expect exciton-phonon scattering and phonon-induced dephasing to exhibit any qualitative differences compared to the monolayer case in terms of  temperature-dependent broadening and diffusion. Therefore, we focus on the impact of an antiferromagnetic exciton-magnon coupling on exciton transport.
The underlying theoretical framework for exciton-magnon coupling in CrSBr bilayers has recently been introduced by Mikhail Glazov and co-workers~\cite{iakovlev2026boltzmann} and is discussed in Supplemental Section VII.  Exciton-magnon interaction in CrSBr bilayers enables tunneling of charge carriers between the layers, which is spin-forbidden in the antiferromagnetic configuration. Magnons can, however, tilt the layer magnetizations and overcome this spin-selection rule, providing an additional tunneling-mediated contribution to the exciton--magnon interaction that is specific to the bilayer geometry. The magnon modes in the two layers hybridize upon diagonalization of the antiferromagnetic magnon Hamiltonian, which enables additional two-magnon processes to become efficient compared to the monolayer case. In particular, two-magnon absorption and emission processes become possible, i.e., terms proportional to $m^2$ and $(m^{\dagger})^2$ enter the bilayer exciton-magnon Hamiltonian. The relevant scattering processes are schematically illustrated in the Feynman diagrams in Fig.~\ref{fig5}(a) with a detailed derivation of the full dephasing provided in Supplemental Section VI. The leading scattering process for the temperature-dependent magnon-induced linewidth in bilayers (Fig.~\ref{fig5}(a)) is the two-magnon exciton-magnon scattering, as is the case also for monolayers. Two-magnon emission is inefficient due to the fact that the initial exciton energy needs to be comparable to twice the magnon energy around the $\Gamma$-point. The qualitative parabolic temperature dependence of the linewidth is retained also for the CrSBr bilayer consistent with Ref.~\cite{iakovlev2026boltzmann}. We note that intralayer spin-flip processes within each layer, analogous to those considered for the ferromagnetic monolayer (cf. Eq. \eqref{2magham}), should also contribute to exciton--magnon interaction in bilayers, but are disregarded here. Such processes are expected to primarily quantitatively enhance the two-magnon exciton-magnon scattering contribution to the magnon-induced dephasing. 

In Fig. ~\ref{fig5}(b), we show the temperature-dependent exciton diffusion coefficient for the CrSBr bilayer including the relevant two-magnon emission, absorption, and scattering processes. Since the overall magnon-induced dephasing increases quadratically with temperature, i.e., similar to the monolayer case, we obtain the same characteristic temperature dependence of the diffusion coefficient for both CrSBr monolayers and bilayers. Furthermore, the anisotropy in the exciton diffusion coefficients is again governed by the anisotropy in translational exciton masses. Overall, while antiferromagnetic ordering introduces additional magnon-assisted scattering channels, it does not qualitatively alter the exciton transport. Instead, the distinctive temperature dependence and strong anisotropy of magnon-limited exciton diffusion remain robust features of CrSBr.

\begin{figure}[t]
    \centering
    \includegraphics[width=\columnwidth]{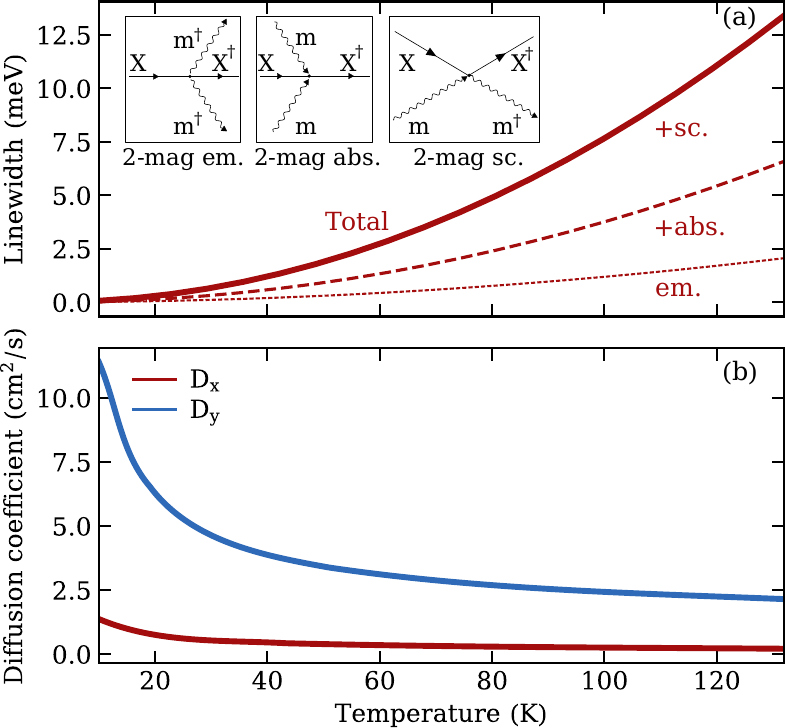}
    \caption{Impact of exciton-magnon interaction on exciton transport in CrSBr bilayers. (a) Exciton linewidth due to exciton-magnon interaction. The relevant two-magnon emission, absorption and scattering processes contributing to the linewidth are schematically shown as Feynman diagrams in the insets. Note that the contributions from two-magnon emission (dashed lines), two-magnon absorption and two-magnon scattering are added up cumulatively. (b) Exciton diffusion coefficient in CrSBr bilayers revealing anisotropic propagation in x- and y-directions (red and blue lines, respectively). }
    \label{fig5}
\end{figure}

%%%%%%%%%%%%%%%%%%%%%%%%%%%%%%%%%%%%%%%%%%%%%%%%%%%%%%%%%%%%%%%%
\section{Conclusions}
%%%%%%%%%%%%%%%%%%%%%%%%%%%%%%%%%%%%%%%%%%%%%%%%%%%%%%%%%%%%%%%%
We have developed a microscopic approach to model exciton--phonon and exciton--magnon interactions and have investigated their impact on exciton transport in the linear regime in mono- and bilayer CrSBr as a prototypical example for a 2D magnetic semiconductor. We show that both scattering mechanisms give rise to strongly anisotropic exciton dephasing and exciton propagation, reflecting the quasi-one-dimensional nature of excitons in anisotropic semiconductors. While exciton--phonon scattering exhibits the characteristic crossover from acoustic- to optical-phonon-dominated dephasing as a function of temperature, exciton--magnon scattering is governed by two-magnon processes, resulting in a distinct quadratic temperature dependence. Our results provide a unified microscopic picture of how lattice and spin excitations govern exciton dynamics in magnetic van der Waals semiconductors and establish a theoretical framework for interpreting future optical spectroscopy and exciton transport experiments.\\

%%%%%%%%%%%%%%%%%%%%%%%%%%%%%%%%%%%%%%%%%%%%%%%%%%%%%%%%%%%%%%%%
\section{Acknowledgments}
%%%%%%%%%%%%%%%%%%%%%%%%%%%%%%%%%%%%%%%%%%%%%%%%%%%%%%%%%%%%%%%%
We thank Mikhail Glazov and Zakhar Iakovlev (Ioffe Institute), Alexey Chernikov (TU Dresden) and Florian Dirnberger (TU Munich) as well as Joshua Thompson (University of Cambridge) for useful discussions. D.E., K.S. and E.M. acknowledge funding by the Deutsche Forschungsgemeinschaft (DFG) via the regular project 571553489 and by the State of Hesse via the LOEWE Priority Programme QuEnergy. T.D. acknowledges financial support from the Deutsche Forschungsgemeinschaft (DFG, German Research Foundation) through Project No. 426726249 (DE 2749/2-1 and DE 2749/2-2). The authors gratefully acknowledge the Gauss Centre for Supercomputing e.V. (www.gauss-centre.eu) for funding this project by providing computing time through the John von Neumann Institute for Computing (NIC) on the GCS Supercomputer JUWELS~\cite{alvarez2021juwels} at Jülich Supercomputing Centre (JSC).

\end{document}

\renewcommand{\theequation}{S\arabic{equation}}
\renewcommand{\thesection}{\Roman{section}}
\preprint{APS/123-QED}

%%%%%%%%%%%%%%%%%%%%%%%%%%%%%%%%%%%%%%%%%%%%%%%%%%%%%%%%%%%%%%%%

\title{Supplemental Material \---  }
\title{\textbf{Supplemental Material} \\ Impact of exciton-phonon and exciton-magnon interaction on transport in anisotropic 2D magnetic semiconductors}
\author{Daniel Erkensten}
\email{*Email: daniel.erkensten@physik.uni-marburg.de}
\affiliation{Department of Physics, Philipps-Universit{\"a}t Marburg, 35037 Marburg, Germany}
\affiliation{mar.quest|Marburg
Center for Quantum Materials and Sustainable Technologies,
35032 Marburg, Germany}
\author{Katarzyna Sadecka}
\affiliation{Department of Physics, Philipps-Universit{\"a}t Marburg, 35037 Marburg, Germany}
\affiliation{mar.quest|Marburg
Center for Quantum Materials and Sustainable Technologies,
35032 Marburg, Germany}
\author{Marie-Christin Heißenbüttel}
\affiliation{
Institute of Solid State Theory, University of Münster, 48149 Münster, Germany}
\author{Thorsten Deilmann}
\affiliation{
Institute of Solid State Theory, University of Münster, 48149 Münster, Germany}
\author{Ermin Malic}
\affiliation{Department of Physics, Philipps-Universit{\"a}t Marburg, 35037 Marburg, Germany}
\affiliation{mar.quest|Marburg
Center for Quantum Materials and Sustainable Technologies,
35032 Marburg, Germany}

\maketitle 
\date{\today}

%%%%%%%%%%%%%%%%%%%%%%%%%%%%%%%%%%%%%%%%%%%%%%%%%%%%%%%%%%%%%%%%

\section{Anisotropic Wannier equation}

The two-dimensional anisotropic Wannier equation reads 
\begin{equation}
(\frac{\hbar^2 k_x^2}{2\mu_x}+\frac{\hbar^2 k_y^2}{2\mu_y})\varphi_{n}(k_x, k_y)-\sum_{\mathbf{q}}W(q_x, q_y)\varphi_n(k_x+q_x, k_y+q_y)=E^b_{n}\varphi_n(k_x, k_y) \ ,
\label{wanniereq}
\end{equation}
where $\mu_i=m_{e,i}m_{h,i}/(m_{e,i}+m_{h,i})$ are reduced exciton masses in the $i=x,y$ directions, corresponding to the $\Gamma-X$ and $\Gamma-Y$ directions for CrSBr, respectively. The carrier masses are obtained directly from the \emph{ab initio}-calculated band structure (Fig. 2(a) in the main text). Furthermore, $\mathbf{k}=(k_x,k_y)$ and $\mathbf{q}=(q_x, q_y)$ are two-dimensional relative momenta, while $\varphi_{n}$ and $E^b_n$ are the exciton wave function and binding energy of the state $n$, respectively. The Coulomb interaction in a CrSBr monolayer reads 
\begin{equation}
W(q_x,q_y)=\frac{V_{\mathbf{q}}}{\varepsilon(q_x, q_y)} \ , 
\end{equation}
where $V_{\mathbf{q}}=\frac{e_0^2}{2\epsilon_0 A |\mathbf{q}|}$ is the bare two-dimensional (2D) Coulomb potential and the dielectric screening is taken to be of the Rytova-Keldysh-form~\cite{keldysh1979coulomb, rytova1967screened}, where  
\begin{equation}
\varepsilon(q_x, q_y)=1+\frac{r_{0,x}q_x^2+r_{0,y}q_y^2}{\epsilon_s|\mathbf{q}|} \ ,
\label{keldysh}
\end{equation}
depends on the screening radii $r_{0,x}$ and $r_{0,y}$ as well as the background dielectric constant $\epsilon_s$. The linearized anisotropic Keldysh screening in Eq.~\eqref{keldysh} was derived by solving the anisotropic Poisson equation~\cite{Semina2025}. The screening radii $r_{0,x}$ and $r_{0,y}$ are obtained by fitting the linearized Keldysh screening in Eq.~\eqref{keldysh} to the fully microscopic dielectric screening calculated using \emph{ab initio} calculations in Ref.~\cite{Klein2023}. We find $r_{0,x}\approx{6.3}$ nm and $r_{0,y}\approx{3}$ nm. Furthermore, in the case of a bilayer, one has to distinguish between intra- and interlayer electron-hole Coulomb potentials~\cite{ovesen2019interlayer, erkensten2022microscopic}, where 
\begin{align}
W_{\mathrm{intra}}(q_x, q_y)=V_{\mathbf{q}}\frac{[1+g_{\mathbf{q}}(1-\mathrm{e}^{-2dq})]}{(1+g_{\mathbf{q}})^2-g^2_{\mathbf{q}}\mathrm{e}^{-2qd}} \ , \label{ekvation1}
\\
W_{\mathrm{inter}}(q_x, q_y)=V_{\mathbf{q}}\frac{\mathrm{e}^{-qd}}{(1+g_{\mathbf{q}})^2-g^2_{\mathbf{q}}\mathrm{e}^{-2qd}} \ , 
\end{align} 
and $g_{\mathbf{q}}=V_{\mathbf{q}}(\alpha_x q_x^2+\alpha_y q_y^2)$ introducing the direction-dependent polarizabilities $\alpha_{i}=2r_{0,i}\epsilon_0 \epsilon_s$ and the CrSBr layer thickness $d\approx{0.8}$ nm. In the limit $d\rightarrow \infty$, the interlayer potential vanishes and $W_{\mathrm{intra}}=W$. In the opposite limit, $d\rightarrow 0$, $W_{\mathrm{inter}}=W_{\mathrm{intra}}=W|_{\alpha_i\rightarrow 2\alpha_i}$, i.e. the polarizabilities and screening radii becomes multiplied by a factor of 2.

The solution of the anisotropic Wannier equation provides microscopic access to the quasi-one dimensional excitons of CrSBr. We note that due to the larger asymmetry in the reduced masses ($\mu_x/\mu_y\approx 8$) compared to the asymmetry in the screening radii ($r_{0,x}/r_{0,y}\approx 2.5$), we find that the quasi-one-dimensional nature of the wave function is predominantly due to the anisotropic electronic band structure of CrSBr. In Table 1, we summarize the relevant material-specific parameters for evaluating exciton binding energies and exciton wave functions in CrSBr mono- and bilayers.  
\begin{table}[h!]
    \centering
    \bgroup
\def\arraystretch{1.5}%  
  \begin{tabular}{c|c|c}
\hline 
    Parameter  & Value & Ref.\\ \hline 
    Screening radius $r_{0,x}$ [nm] & 6.3 &~\cite{Klein2023} \\ \hline  
    Screening radius $r_{0,y}$ [nm] & 3 &~\cite{Klein2023} \\ \hline  
    Electron mass $m_{e,x}$ ($m_{e,y}$) [$m_0$]& 4.96 (0.45) & This work \\ \hline 
    Hole mass $m_{h,x}$ ($m_{h,y}$) [$m_0$] & 2.84 (0.45) & This work\\ \hline 
    Dielectric background constant $\epsilon_s$ & 4.5 &~\cite{PhysRev.146.543}  \\ \hline 
    Layer thickness $d$ [nm] & 0.8 & ~\cite{Wilson2021}
\end{tabular}
\egroup
    \caption{Material-specific parameters used as input for the anisotropic Wannier equation. The carrier masses are taken along the $\Gamma-X$ and $\Gamma-Y$ directions in the vicinity of the $\Gamma$-point and extracted from the \emph{ab initio} band structure of a CrSBr monolayer. }
    \label{parametertabell_magnon}
\end{table}

%%%%%%%%%%%%%%%%%%%%%%%%%%%%%%%%%%%%%%%%%%%%%%%%%%%%%%%%%%%%%%%%
\section{Excitonic wave function from Bethe-Salpeter equation}
The calculation of the exciton wave function from first principles
builds up on the solution of the Bethe-Salpeter Equation (BSE)  \cite{PhysRevB.62.4927, RevModPhys.74.601}.
In the electron-hole basis the matrix elements (for exciton momentum $Q=0$) are calculated as
\begin{equation}\label{eq:BSEmatrix}
  (E_{ck}-E_{vk}) A^S_{vck} + \sum_{v'c'k'} K^{AA}_{vck,v'c'k'}(\Omega_S) A^S_{v'c'k'} = \Omega_S A^S_{vck} .
\end{equation}
and the diagonalization leads to the exciton energies $\Omega_S$ and their amplitudes $A^S_{vck}$.
In reciprocal space, the exciton wave function
is calculated via $\sum_{{v}{c}} |A^{S}_{vck}|^2$, while
in real space it is evaluated by
\begin{equation}\label{eq:exwf}
\Phi^{S}(x_h,x_e) = \sum_{{v}{c}} A^{S}_{vc} \phi_{v}^\ast(x_h) \phi_{c}(x_e) .
\end{equation}

\begin{figure}[b!]
    \centering
    \includegraphics[width=0.5\linewidth]{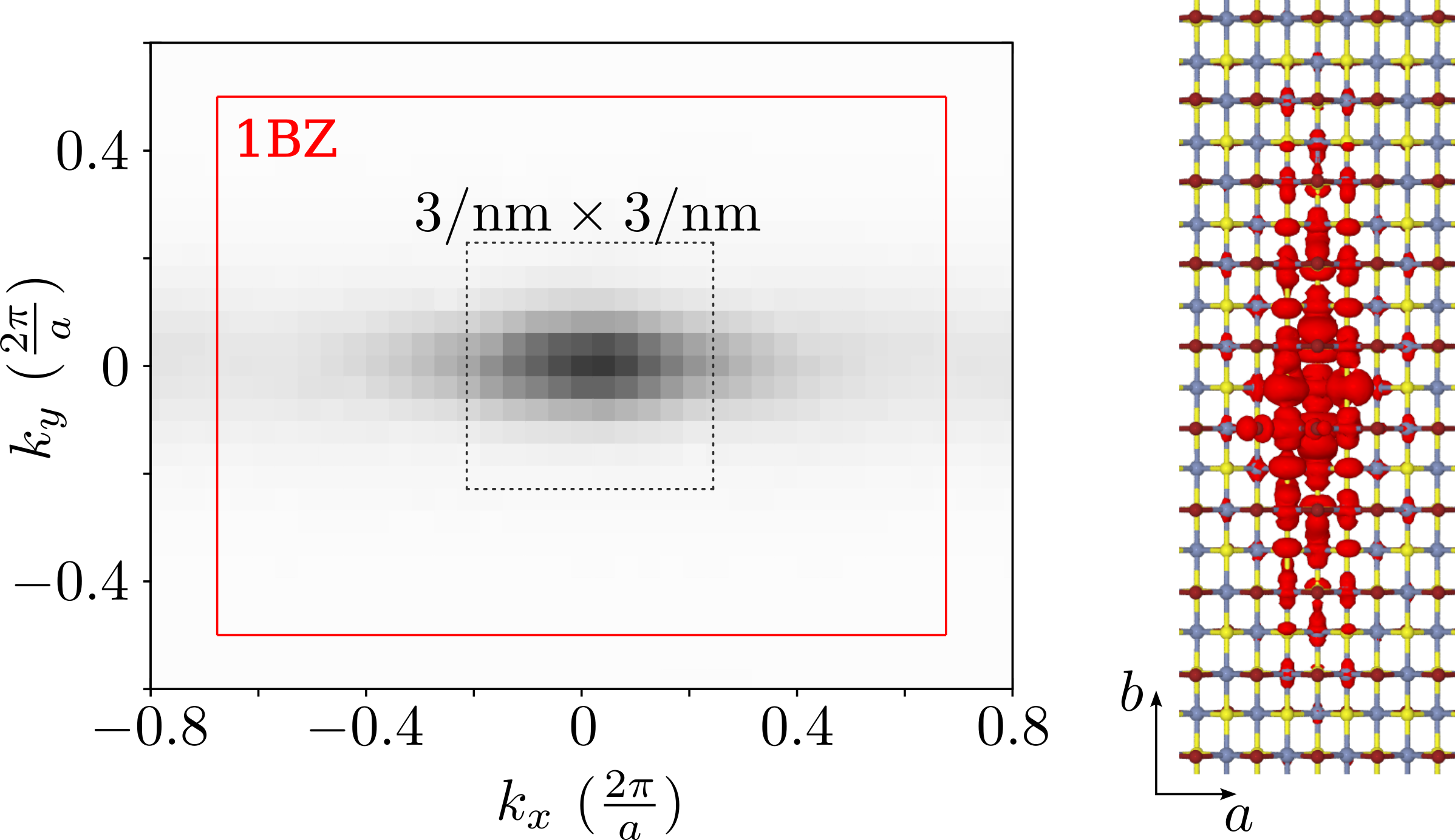}
    \caption{1$s$ exciton wave function in monolayer CrSBr as a function of momentum obtained from BSE. The rectangular Brillouin zone has been indicated with solid red lines, while the dashed black lines represent the region considered in the main text. Right panel: real-space probability of electron with hole fixed close to the Cr atoms. }
    \label{bsesol}
\end{figure}

In Fig. \ref{bsesol} we show the wave function of the lowest-energy exciton in a CrSBr monolayer.
The red box denotes the 1st Brillouin zone with the $32\times 24$ k-points evaluated using the BSE. For comparison, the dashed box marks the region plotted in Fig. 2b in the main text. 
The right panel shows the probability of the electron, when the hole is fixed close to the Cr atoms, indicating a highly anisotropic and spatially extended exciton along the b-axis - indicating a Wannier-type exciton character in CrSBr monolayers.
%%%%%%%%%%%%%%%%%%%%%%%%%%%%%%%%%%%%%%%%%%%%%%%%%%%%%%%%%%%%%%%%
\newpage 
\section{Ab initio calculations of electrons and phonons}

Our ab initio calculations have been carried out
within density functional theory (DFT) and many-body perturbation theory (MBPT).
We take a CrSBr unit cell size of $4.572$\AA$\times 3.512$\AA~\cite{Heissenbuttel2025}.
The atomic positions are relaxed until all forces are below $10^{-4}$\,Ry/$a_B$.
In the following we briefly recapitulate our computation methods~\cite{Heissenbuttel2025}.
As the system is ferromagnetic, both spin-polarization and spin-orbit coupling
have been taken fully into account,
resulting in the noncollinear formalism of the spin density approach
\cite{stark2011magnetic}.
Starting from the generalized gradient approximation (GGA) 
we perform a full spinor $GW$ calculation approximating MBPT~\cite{PhysRevB.101.085130}.
The solution of the Hamiltonian
\begin{equation}
  H^\text{QP} = H^\text{DFT} + iGW - V_\text{xc},
\end{equation}
where $G$ is the Green function, $W$ the screened Coulomb interaction and $V_\text{xc}$
the exchange-correction potential from the DFT,
leads to the diagonal quasiparticle energies
\begin{equation}
  \tilde{E}^\text{QP}_{n\vec{k}} = E^\text{DFT}_{n\vec{k}} + \langle{\psi^\text{DFT}_{n\vec{k}}| \left(\Sigma(E^\text{QP}_{n\vec{k}}) - V_\text{xc}\right) | \psi^\text{DFT}_{n\vec{k}}}\rangle  .
\end{equation}
However, we find that such one-shot $GW$ is not sufficient for CrSBr
and include non-diagonal elements such that
\begin{equation}
  \sum_{n'} H^\text{QP}_{nn'}(\vec{k},E^\text{QP}_{m\vec{k}}) D^{n'}_{m\vec{k}} = E^\text{QP}_{m\vec{k}} D^{n}_{m\vec{k}} \ , 
\end{equation}
leads to the final band structure energies
and modified quasiparticle wavefunctions $\psi^\text{QP}_{n\vec{k}} = \sum_{n} D^{n}_{m\vec{k}} \psi^\text{DFT}_{n\vec{k}}$.
For our quasiparticle calculations for the CrSBr monolayer,
we employ a hybrid basis set of Gaussian orbitals (decay constants from 0.14 to 12.5\,$a_B^{-2}$)
and plane waves (energy cutoff 1.5\,Ry)
to represent of all two-point quantities.
We sample the first Brillouin with $32\times 24$ $\vec{k}$ points.

To evaluate the phonon dispersion in the system~\cite{mann2024electron},
we evaluate the force constants by displacing each atom in the unit cell
and diagonalize the corresponding dynamic matrix.
We find a good convergence using a $6\times 6$ supercell when using DFT-LDA.
Note that the tiny influence of spin-orbit interactions on the phonons has been neglected
while the spin-polarization has to be taken into account.
To avoid numerical instabilities of the force constants,
we restrict them to interactions to atoms closer than $7.4$\,\AA,
i.e. take at least the 5-th next neighbours into account.
% Potentially add tables with velocities and optical phonon energies 
\section{Magnon dispersion}
\subsection{Ferromagnetic monolayer}
In order to derive the magnon dispersion of the ferromagnetic CrSBr monolayer, we start from the Heisenberg spin Hamiltonian supplemented with lattice anisotropy terms
\begin{equation}
H^{(\mathrm{m})}=-\frac{1}{2}\sum_{\langle i,j\rangle}J_{ij}\mathbf{S}_i\cdot\mathbf{S}_j+\sum_{i}(A_x S_{i, x}^2+A_{z}S_{i,z}^2) \ , 
\end{equation}
where the first term is the regular Heisenberg Hamiltonian including the ferromagnetic exchange interaction between nearest-neighbors (NN), $J_{<ij>}>0$, and the second term is due to lattice anisotropy. The spin operators $\mathbf{S}_i$ describe localized magnetic moments at site $i$. Here, we choose $A_x, A_z>0$ such that the y-axis becomes the easy axis (b-axis). Hence, the corresponding Holstein-Primakoff transformation that relates the spin operators to magnon operators $a^{(\dagger)}_i$ reads~\cite{holstein1940field}
\begin{equation}
\begin{aligned}
S_{j,y}&=s-a^{\dagger}_j a_j \ , \\
S_{j,x}&=\sqrt{\frac{s}{2}}i(a^{\dagger}_j-a_j) \ , \\
S_{j,z}&=\sqrt{\frac{s}{2}}(a^{\dagger}_j+a_j) \ ,  
\end{aligned}
\label{hp}
\end{equation}
which holds in the limit of low temperatures and small magnon populations $\langle a^{\dagger}_i a_i\rangle$. The spin operator $S_y$ is represented in terms of the Pauli matrix $\sigma_z=\mathrm{diag}(1,-1)$ (the other components are obtained from cyclic permutation). Furthermore, the spin $s=\frac{3}{2}$ for the localized magnetic moments stems from Cr$^{3+}$ ions in CrSBr. Assuming that the spins in the Heisenberg Hamiltonian interact only via NN exchange interactions and introducing the Fourier transform $a_i=\frac{1}{\sqrt{N}}\sum_{\mathbf{k}}\mathrm{e}^{i\mathbf{k}\cdot\mathbf{r}_i}a_{\mathbf{k}}$, the spin Hamiltonian is rewritten in terms of the magnon operators as 
\begin{equation}
H^{(\mathrm{m})}=\sum_{\mathbf{k}}\big(B(\mathbf{k})+s(A_x+A_z)\big)a^{\dagger}_{\mathbf{k}}a_{\mathbf{k}}+\frac{s}{2}(A_z-A_x)\sum_{\mathbf{k}}(a^{\dagger}_{\mathbf{k}}a^{\dagger}_{\mathbf{-k}}+a_{\mathbf{k}}a_{\mathbf{-k}}) \ , 
\label{offdiag}
\end{equation}
where $B(\mathbf{k})=2s\sum_{n=1}^3 J_n (\gamma_{n}(\mathbf{k})-\lambda_n)$ with $\gamma_n(\mathbf{k})=\sum_{\boldsymbol{\delta}_n}\cos(\boldsymbol{\delta}_n\cdot\mathbf{k})$ and $\lambda_n$ is the coordination number (the number of equivalent NNs for each $J_n$). For the rectangular lattice of CrSBr, we find $\boldsymbol{\delta}_1\in \{(a_0,0)^T, (-a_0,0)^T\}$, $\boldsymbol{\delta}_2\in \{(a_0/2, b_0/2)^T, (-a_0/2, b_0/2)^T,(a_0/2, -b_0/2)^T,(-a_0/2, -b_0/2)^T\}$ and $\boldsymbol{\delta}_3\in \{(0, b_0)^T, (0,-b_0)^T\}$, with $a_0$ and $b_0$ being the in-plane lattice constants of CrSBr. Note that due to the lattice anisotropies $A_z$ and $A_x$, the magnon Hamiltonian is not diagonal. In order to diagonalize the Hamiltonian, we rewrite it in the form $H^{(\mathrm{m})}=\frac{1}{2}\sum_{\mathbf{k}}\psi^{\dagger}_{\mathbf{k}}\mathcal{H}^{(\mathrm{m})} \psi_{\mathbf{k}}$, where the spinor $\psi_{\mathbf{k}}=\begin{pmatrix}a_{\mathbf{k}}\\
a^{\dagger}_{-\mathbf{k}}
\end{pmatrix}$ and 
\begin{equation}
\mathcal{H}^{(\mathrm{m})}=\begin{pmatrix}
B(\mathbf{k})+s(A_x+A_z) & s(A_z-A_x) \\
s(A_z-A_x) & B(\mathbf{k})+s(A_x+A_z)  
\end{pmatrix} \ . 
\end{equation}
Since the excitations are bosonic, the relevant Hamiltonian matrix is $\bar{\mathcal{H}}^{(\mathrm{m})}=\sigma_z \mathcal{H}^{(\mathrm{m})}$, where $\sigma_z=\mathrm{diag}(1,-1)$. Diagonalization of this Hamiltonian results in the eigenenergies 
\begin{equation}
\hbar\Omega^{(\mathrm{m})}_{\mathbf{k}}=\sqrt{[B(\mathbf{k})+s(A_x+A_z)]^2-s^2(A_z-A_x)^2} \ . 
\end{equation}
Importantly, we find that the energy dispersion has a small gap $\hbar\Omega^{(\mathrm{m})}_{\Gamma}\equiv \Delta=2s\sqrt{A_x A_z}\approx{0.1}$ meV around the $\Gamma$-point. Here, we employed anisotropy parameters $A_x=14$ $\mu$eV and $A_z=58$ $\mu$eV as obtained from Ref.~\cite{Bae2022}. In addition, for small $k$, the dispersion can be approximated as parabolic in momentum, i.e. 
\begin{equation}
\hbar\Omega^{(\mathrm{m})}_{\mathbf{k}}\approx \Delta+\frac{\hbar^2 k_x^2}{2M_x^{(\mathrm{m})}}+\frac{\hbar^2 k_y^2}{2M_y^{(\mathrm{m})}} \ , 
\label{parabolic}
\end{equation}
where the magnon masses depends on the NN exchange couplings as $M_x^{(\mathrm{m})}=\frac{\hbar^2}{a_0^2 (2J_1+J_2)}$ and $M_x^{(\mathrm{m})}=\frac{\hbar^2}{b_0^2 (2J_3+J_2)}$. With $J_1=1.9$ meV, $J_2=3.38$ meV and $J_3=1.67$ meV~\cite{Bo_2023}, we find $M_x^{(\mathrm{m})}\approx 57 m_0$ and $M_y^{(\mathrm{m})}\approx 34 m_0$. Note that the derivation outlined above, only assumed a single magnetic sublattice per unit cell, whereas CrSBr has two sublattices per magnetic unit cell. The high-energy magnon mode is therefore obtained by zone-folding. Finally, the magnon Hamiltonian in Eq.~\eqref{offdiag} is rewritten in the diagonal form $H^{(\mathrm{m})}=\sum_{\mathbf{k}} \hbar\Omega^{(\mathrm{m})}_{\mathbf{k}}m^{\dagger}_{\mathbf{k}}m_{\mathbf{k}}$ via the Bogoliubov transformation
\begin{equation}
\begin{aligned}
a_{\mathbf{k}}&=u_{\mathbf{k}}m_{\mathbf{k}}+v_{\mathbf{k}}m^{\dagger}_{-\mathbf{k}} \ , \\
a_{-\mathbf{k}}&=u_{\mathbf{k}}m_{-\mathbf{k}}+v_{\mathbf{k}}m^{\dagger}_{\mathbf{k}} \ , 
\end{aligned}
\end{equation}
where the Bogoliubov coefficients $u^2_{\mathbf{k}}=\frac{B(\mathbf{k})+s(A_x-A_z)+\hbar\Omega^{\mathrm{m}}_{\mathbf{k}}}{2\hbar\Omega^{(\mathrm{m})}_{\mathbf{k}}}$ and $v^2_{\mathbf{k}}=\frac{B(\mathbf{k})+s(A_x-A_z)-\hbar\Omega^{(\mathrm{m})}_{\mathbf{k}}}{2\hbar\Omega^{(\mathrm{m})}_{\mathbf{k}}}$. Since the lattice anisotropies are of the order of $\mu$eVs and the exchange couplings of the order of meVs, we approximate $u_{\mathbf{k}}^2\approx 1$ and $v^2_{\mathbf{k}}\approx 0$. Within this approximation it follows for the magnon operators $m^{(\dagger)}\approx a^{(\dagger)}$. 

\subsection{Antiferromagnetic bilayer}
In an antiferromagnetic CrSBr bilayer, one has to introduce two sets of magnon operators, $a^{(\dagger)}$ and $b^{(\dagger)}$, acting on each layer with the spins aligned in opposite directions~\cite{iakovlev2026boltzmann}. Then, one has to perform a generalized Bogoliubov transformation, resulting in new magnon operators $m^{(\dagger)}_{1}$ and $m^{(\dagger)}_2$ and leading to two low-energy magnon modes with the energy 
\begin{equation}
\hbar \Omega^{(\mathrm{m})}_{\pm, \mathbf{k}}=\sqrt{[B(\mathbf{k})+s(A_x+A_z)+s J_{\mathrm{int}}]^2-s^2[(A_z-A_x)\mp J_{\mathrm{int}}]^2} \ , 
\end{equation}
with the interlayer exchange coupling $J_{\mathrm{int}}$. Note that since the interlayer exchange coupling is weak and of the order $\sim 1$ $\mu$eV, the two low-energy modes become approximately degenerate and their energy coincides with that of the ferromagnetic monolayer. The diagonal form of the Hamiltonian reads $H^{(\mathrm{BL})}_{M}=\sum_{\mathbf{k},i}\hbar \Omega^{(\mathrm{m})}_{i, \mathbf{k}}m^{\dagger}_{i,\mathbf{k}}m_{i,\mathbf{k}}$ with $\hbar \Omega^{(\mathrm{m})}_{1, \mathbf{k}}=\hbar \Omega^{(\mathrm{m})}_{+, \mathbf{k}}$ and $\hbar \Omega^{(\mathrm{m})}_{2, \mathbf{k}}=\hbar \Omega^{(\mathrm{m})}_{-, \mathbf{k}}$. The new magnon operators are related to $a^{(\dagger)}$ and $b^{(\dagger)}$ according to~\cite{iakovlev2026boltzmann}
\begin{equation}
\begin{pmatrix}
a_{\mathbf{k}} \\ 
b_{\mathbf{k}} \\
a^{\dagger}_{-\mathbf{k}}\\
b^{\dagger}_{-\mathbf{k}}
\end{pmatrix}=\begin{pmatrix}
X^{++}_{\mathbf{k}} & -X^{-+}_{\mathbf{k}}  &  -X^{+-}_{\mathbf{k}} &-X^{--}_{\mathbf{k}} \\
X^{++}_{\mathbf{k}} & X^{-+}_{\mathbf{k}}  &  -X^{+-}_{\mathbf{k}} &X^{--}_{\mathbf{k}} \\
-X^{+-}_{\mathbf{k}} & -X^{--}_{\mathbf{k}}  &  X^{++}_{\mathbf{k}} &-X^{-+}_{\mathbf{k}} \\
-X^{+-}_{\mathbf{k}} & X^{--}_{\mathbf{k}}  &  X^{++}_{\mathbf{k}} &X^{-+}_{\mathbf{k}}
\end{pmatrix}\begin{pmatrix}
m_{1,\mathbf{k}} \\ 
m_{2,\mathbf{k}} \\
m^{\dagger}_{1,-\mathbf{k}}\\
m^{\dagger}_{2,-\mathbf{k}}
 \end{pmatrix} \ , 
 \label{bogolon}
\end{equation}
and $X^{\pm \pm'}_{\mathbf{k}}=\frac{1}{2}\sqrt{\frac{B(\mathbf{k})+s(A_x+A_z)+s J_{\mathrm{int}}}{\hbar\Omega^{(\mathrm{m})}_{\pm,\mathbf{k}}}\pm' 1 }$. The relevant material-specific parameters used for evaluating the magnon dispersion in the CrSBr mono- and bilayers are reported in Table II. 
\begin{table}[h!]
    \centering
    \bgroup
\def\arraystretch{1.5}%  
  \begin{tabular}{c|c|c}
\hline 
    Parameter  & Value & Ref.\\ \hline 
    Lattice constant $a_0$ [nm] & 0.35 &~\cite{Bo_2023} \\ \hline  
    Lattice constant $b_0$ [nm] & 0.47 &~\cite{Bo_2023} \\ \hline 
    Lattice anisotropy $A_x$ [$\mu$eV] & 14 &~\cite{Bae2022} \\ \hline  
    Lattice anisotropy $A_z$ [$\mu$eV] & 58 &~\cite{Bae2022} \\ \hline  
    NN exchange coupling $J_1$ [meV] & 1.9 &~\cite{Bo_2023}\\ \hline 
    NN exchange coupling $J_2$ [meV] & 3.38 &~\cite{Bo_2023}\\ \hline 
    NN exchange coupling $J_3$ [meV] & 1.67 & 
   ~\cite{Bo_2023}\\ \hline
    Interlayer exchange coupling $J_{\mathrm{int}}$ [$\mu$eV]  & 6 & 
   ~\cite{Bo_2023}
\end{tabular}
\egroup
    \caption{Material-specific parameters used for evaluation of magnon dispersions in mono- and bilayer CrSBr.}
    \label{parametertabell_magnon}
\end{table}

%%%%%%%%%%%%%%%%%%%%%%%%%%%%%%%%%%%%%%%%%%%%%%%%%%%%%%%%%%%%%%%%

%\newpage 
\section{Deformation potentials and exciton-phonon scattering}
The phonon-induced dephasing in two-dimensional semiconductors reads~\cite{selig2016excitonic}
\begin{equation}
\gamma_{\mathrm{ph}}(\mathbf{Q})=\frac{\pi}{\mathcal{S}}\sum_{j, \mathbf{q}, \pm}|G^{(\mathrm{ph})}_{j, \mathbf{q}}|^2 (n^{(\mathrm{ph})}_{j, \mathbf{q}}+\frac{1}{2}\pm \frac{1}{2})\delta(\epsilon^{\pm}_{j, \mathbf{Q},\mathbf{q}}) \ , 
\label{gammaph}
\end{equation}
where $\epsilon^{\pm}_{j, \mathbf{Q},\mathbf{q}}=E_{\mathbf{Q+q}}-E_{\mathbf{Q}}\pm \hbar\Omega^{(\mathrm{ph})}_{j, \mathbf{q}}$ and $n^{(\mathrm{ph})}_{j, \mathbf{q}}$ is taken as a thermalized Bose-Einstein distribution and $\mathcal{S}$ is the crystal area. Furthermore, the exciton-phonon coupling $G$ is given by 
\begin{equation}
G^{(\mathrm{ph})}_{j, \mathbf{q}}=g^c_{j, \mathbf{q}}F(\boldsymbol{\beta}\cdot\mathbf{q})-g^{v}_{j, \mathbf{q}}F(-\boldsymbol{\alpha}\cdot\mathbf{q}) \ , 
\end{equation}
introducing the electron-phonon coupling matrix element $g^{\lambda}_{j, \mathbf{q}}$, $\lambda=c,v$ and the exciton form factor $F(\mathbf{q})=\sum_{\mathbf{k}}\varphi^*_{1s, \mathbf{k+q}}\varphi_{1s,\mathbf{k}}$. Here, we focus on the coupling between phonons and the energetically lowest $n=1s$ excitons with the exciton wave function being obtained from solving the anisotropic Wannier equation (Eq.~\eqref{wanniereq}).  In the main text, we have omitted the crystal area $\mathcal{S}$ in the expression for the dephasing, as the dephasing becomes independent of the crystal area when transforming the momentum sum over $\mathbf{q}$ into an integral. 

The electron-phonon coupling matrix elements are treated within the deformation potential approximation~\cite{PhysRevB.90.045422}:
\begin{equation}
g^{\lambda}_{j, \mathbf{q}}=\sqrt{\frac{\hbar}{2\rho \Omega_{j, \mathbf{q}}}}D^{\lambda}_{j,\mathbf{q}} \ , 
\end{equation}
where $\rho$ is the 2D surface mass density, $D^{\lambda}_{j,\mathbf{q}}$ is the deformation potential for the electronic band $\lambda$ and phonon mode $j$ and $\Omega_{j, \mathbf{q}}$ is the phonon frequency. Note that the deformation potentials in valence and conduction bands have opposite signs, i.e. the bands shift in opposite directions under strain. For long-wavelength acoustic phonons it holds that $D^{\lambda}_{ac, \mathbf{q}}=\tilde{D}^{\lambda}_{ac}|\mathbf{q}|$, whereas for optical phonons, $D^{\lambda}_{op, \mathbf{q}}=\tilde{D}^{\lambda}_{op}$~\cite{PhysRevB.90.045422}. The relevant deformation potentials for electron-phonon coupling in TMDs at relevant high-symmetry points as obtained from density functional perturbation
theory are reported in Ref.~\cite{PhysRevB.90.045422}. To the best of our knowledge, the deformation potentials describing electron–phonon scattering in CrSBr have not yet been reported. Since we do not have microscopic access to these potentials, we extract them by fitting the microscopically calculated phonon-induced dephasing at $Q=0$ to the recently measured temperature-dependent linewidth for CrSBr~\cite{Dirnberger2026}. In order to reduce the number of fitting parameters, we average over the relevant transversal acoustic (TA) and longitudinal acoustic (LA) modes with a sound velocity given by the arithmetic average of the TA and LA velocities (including averaging over $\Gamma-X$ and $\Gamma-Y$ directions). In addition, we average the 15 optical modes into one optical mode with the optical phonon energy $E_{\mathrm{avg}, op}=24.8$ meV using the Einstein approximation. In addition, we consider the electron and hole deformation potentials to be equal. Furthermore, we disregard the impact of anisotropy on the electron-phonon coupling and consider the coupling as isotropic in momentum. Hence, the phonon energies are given by $\hbar\Omega_{ac, \mathbf{q}}=\hbar v_{\mathrm{avg}, ac} |\mathbf{q}|$, where $v_{\mathrm{avg}, ac}$ is the speed of sound, and $\hbar\Omega_{op, \mathbf{q}}=E_{\mathrm{avg}, op}$. The deformation potential parameters obtained from the fitting procedure as well as the averaged sound velocity and optical phonon energy are provided in Table III. We note that the obtained phonon parameters are similar to those of MoSe$_2$ monolayers, in which $v_{ac, \Gamma}=0.0041$ nm/fs, $\bar{D}_{ac, K\rightarrow K}\approx {2-2.4}$ eV, whereas the relevant optical phonon energies $E_{\mathrm{op}, \Gamma}\approx 30-35$ meV and optical deformation coupling $\bar{D}_{op, K\rightarrow K}\approx 50$ eV/nm are slightly larger in MoSe$_2$ than for CrSBr~\cite{PhysRevB.90.045422}. 
\begin{table}[h!]
    \centering
    \bgroup
\def\arraystretch{1.5}%  
  \begin{tabular}{c|c}
\hline 
    Parameter  & Value  \\ \hline  
    Acoustic sound velocity (averaged) $v_{\mathrm{avg}, ac}$ [nm/fs] & 0.0035  \\ \hline  
    Optical phonon energy (averaged) $E_{\mathrm{avg},op}$ [meV]  & 24.8\\ \hline 
    Acoustic deformation coupling $|\bar{D}_{ac}|$ [eV] & 2  \\ \hline  
    Optical deformation coupling $|\bar{D}_{\mathrm{op}}|$ [eV/nm] & 44  \\ \hline  
\end{tabular}
\egroup
    \caption{Material-specific parameters used for evaluation of exciton-phonon scattering around the $\Gamma$-point in CrSBr. }
    \label{parametertabell_magnon}
\end{table}

%%%%%%%%%%%%%%%%%%%%%%%%%%%%%%%%%%%%%%%%%%%%%%%%%%%%%%%%%%%%%%%%

\section{Exciton-magnon interaction in ferromagnetic monolayers}
In this section, we perform a derivation of the exciton-magnon Hamiltonian and the associated magnon-induced dephasing for a ferromagnetic magnetic semiconductor.  
The starting point for the interactions between excitons and magnon in a ferromagnetic semiconductor is the carrier-magnon interaction, in particular the interaction between itinerant electrons and localized magnetic moments~\cite{PhysRevB.1.4474, Jo2021} 
\begin{equation}
H_{\mathrm{c-m}}=-\sum_{i, \lambda} \tilde{I}_{\lambda}\mathbf{S}_i\cdot\mathbf{s}^{\lambda}_i \ , 
\label{sdinteraction}
\end{equation}
where
\begin{equation}
\mathbf{s}^{\lambda}_i=\frac{\hbar}{2N}\sum_{\mathbf{k}, \mathbf{k}'}\sum_{\sigma, \sigma'}\mathrm{e}^{i(\mathbf{k}-\mathbf{k}')\cdot\mathbf{r}_i}\lambda^{\dagger}_{\sigma\mathbf{k}}\vec{\boldsymbol{\sigma}}_{\sigma\sigma'}\lambda_{\sigma' \mathbf{k}'} \ , 
\end{equation}
is the spin density operator of the electrons, $\mathbf{r}_i$ denotes the position of site $i$ and $\lambda=c,v$ is the band index. The operator $\lambda^{\dagger}_{\sigma\mathbf{k}}$ ($\lambda_{\sigma\mathbf{k}}$) creates (annihilates) an electron with spin $\sigma$ and momentum $\mathbf{k}$ in band $\lambda$. Furthermore, $\vec{\boldsymbol{\sigma}}=(\sigma_x, \sigma_y, \sigma_z)$ is a vector of 2 by 2 Pauli matrices. 
By performing a Holstein-Primakoff transformation of $\mathbf{S}$ (cf. Eq~\eqref{hp}) , we obtain the carrier-magnon Hamiltonian: 
\begin{align*}
H_{\mathrm{c-m}}&=-\frac{\sqrt{s}}{\sqrt{2N}}\sum_{\mathbf{k}, \mathbf{q}, \lambda,j }I_{\lambda}[m^{\dagger}_{j,\mathbf{q}}\lambda^{\dagger}_{\downarrow \mathbf{k}-\mathbf{q}}\lambda_{\uparrow \mathbf{k}}+m_{j,\mathbf{q}}\lambda^{\dagger}_{\uparrow \mathbf{k}+\mathbf{q}}\lambda_{\downarrow \mathbf{k}}]\\
&+\frac{1}{2N}\sum_{\mathbf{k}, \mathbf{k}', \mathbf{q}, \lambda,j}I_{\lambda}m^{\dagger}_{j,\mathbf{q}}m_{j,\mathbf{q}-\mathbf{k}'+\mathbf{k}}(\lambda^{\dagger}_{\downarrow \mathbf{k}}\lambda_{\downarrow \mathbf{k}'}-\lambda^{\dagger}_{\uparrow \mathbf{k}}\lambda_{\uparrow \mathbf{k}'}) \\
& -\frac{s}{2}\sum_{\lambda, \mathbf{k}}I_{\lambda}(\lambda^{\dagger}_{\downarrow \mathbf{k}}\lambda_{\downarrow \mathbf{k}}-\lambda^{\dagger}_{\uparrow \mathbf{k}}\lambda_{\uparrow \mathbf{k}})] \ , 
\end{align*}
redefining the carrier-magnon coupling as $I_{\lambda}=\hbar\tilde{I}_{\lambda}$. We note that the interaction contains spin-conserving two-magnon processes (second and third line) and spin-flip one-magnon processes (first line). The third and fourth terms in the second line constitute a two-magnon interaction, being second order in magnon operators. The final two terms renormalize the single-particle carrier energies of the spin-split bands. In fact, 
the energy renormalization due to the hole-magnon coupling has been previously measured for the ferromagnetic semimetal EuCd$_2$As$_2$ by means of ARPES measurements across the Curie temperature~\cite{Jo2021}. Temperature-dependent ARPES measurements could hence be used to extract the hole-magnon coupling, $I_h$.   

In order to get microscopic access to the exciton-magnon interaction, we make use of the pair operator expansion to transfer electronic quantities to excitonic quantities~\cite{katsch2018theory, erkensten2021exciton}
\begin{equation}
c^{\dagger}_{i}c_j\approx \sum_{m}P^{\dagger}_{i,m}P_{j, m} \ ,
\end{equation}
with the pair operator $P^{\dagger}_{i,j}=c^{\dagger}_i v_j$. By expressing the pair operator in the exciton basis, relating the pair operator to the exciton operators $X$ and $X^{\dagger}$ and exciton wave function $\varphi_{n}$~\cite{katsch2018theory}, the exciton-magnon interaction decomposes into one-magnon and two-magnon contributions to the exciton-magnon scattering. The one-magnon exciton-magnon Hamiltonian reads 
\begin{equation}
H^{(1)}_{\mathrm{x-m}}=\sum_{\mathbf{Q}, \mathbf{q}, n,l, \boldsymbol{\sigma}, \boldsymbol{\sigma'},j}M^{nl}_{\boldsymbol{\sigma\sigma'}}(\mathbf{q}) m^{\dagger}_{j,\mathbf{q}}X^{\dagger \boldsymbol{\sigma}}_{n, \mathbf{Q}-\mathbf{q}}X^{\boldsymbol{\sigma'}}_{l,\mathbf{Q}}+\mathrm{h.c.} \ , 
\end{equation}
where 
\begin{align*}
M^{nl}_{\boldsymbol{\sigma}\boldsymbol{\sigma'}}(\mathbf{q})=\frac{\sqrt{s}}{\sqrt{2N}}[I_h F_{nl}(-\boldsymbol{\alpha}^{\boldsymbol{\sigma}}\cdot\mathbf{q})\delta_{\sigma_e, \sigma'_e}\delta_{\sigma_h, 1-\sigma'_h}-I_e F_{nl}(\boldsymbol{\beta}^{\boldsymbol{\sigma}}\cdot\mathbf{q})\delta_{\sigma_e, 1-\sigma'_e}\delta_{\sigma_h, \sigma'_h}] \ .
\end{align*}
Here, we introduced the compound spin index $\boldsymbol{\sigma}=(\sigma_e, \sigma_h)$ and the form factor $F_{nl}(\mathbf{q})=\sum_{\mathbf{k}}\varphi^*_{n,\mathbf{k+q}}\varphi_{l, \mathbf{k}}$. Furthermore, $n$ and $l$ enumerate the exciton state, and the carrier mass ratios $\boldsymbol{\beta}^{\boldsymbol{\sigma}}=1-\boldsymbol{\alpha}^{\boldsymbol{\sigma}}=(m^{\sigma_e}_{e,x}/(m^{\sigma_e}_{e,x}+m^{\sigma_h}_{e,x}),m^{\sigma_e}_{e,y}/(m^{\sigma_e}_{e,y}+m^{\sigma_h}_{e,y}))$. Importantly, the exciton-magnon coupling induces either a spin-flip of the electron ($\propto I_e$) or hole ($\propto I_h$). For the two-magnon processes, we obtain the Hamiltonian 
\begin{align*}
H_{\mathrm{x-m}}^{(2)}&=\frac{1}{2N}\sum_{\mathbf{Q}, \mathbf{q}, \mathbf{q}', \sigma, n,l,j}m^{\dagger}_{j,\mathbf{q}}m_{j,\mathbf{q+q'}}\bigg(I_e[X^{\dagger \downarrow \sigma}_{n, \mathbf{Q}}X^{\downarrow\sigma}_{l,\mathbf{Q}-\mathbf{q}'}F_{nl}(\boldsymbol{\beta}^{\downarrow \sigma}\cdot\mathbf{q}')-X^{\dagger \uparrow \sigma}_{n, \mathbf{Q}}X^{\uparrow\sigma}_{l,\mathbf{Q}-\mathbf{q}'}F_{nl}(\boldsymbol{\beta}^{\uparrow \sigma}\cdot\mathbf{q}')]\\
& -I_h[X^{\dagger \sigma\downarrow}_{n, \mathbf{Q}}X^{\sigma\downarrow}_{l,\mathbf{Q}-\mathbf{q}'}F_{nl}(-\boldsymbol{\alpha}^{\sigma\downarrow}\cdot\mathbf{q}')-X^{\dagger \sigma\uparrow}_{n, \mathbf{Q}}X^{\sigma\uparrow}_{l,\mathbf{Q}-\mathbf{q}'}F_{nl}(-\boldsymbol{\alpha}^{\sigma\uparrow}\cdot\mathbf{q}')]\bigg) \ , 
\end{align*}
which we can be rewritten as  
\begin{align*}
H^{(2)}_{\mathrm{x-m}}=\sum_{\mathbf{Q}, \mathbf{q}, \mathbf{q}', \boldsymbol{\sigma}, \boldsymbol{\sigma'}n,l,j}G_{\boldsymbol{\sigma}\boldsymbol{\sigma}'}^{nl}(\mathbf{q}')m^{\dagger}_{j,\mathbf{q}}m_{j,\mathbf{q+q'}}X^{\dagger \boldsymbol{\sigma}}_{n, \mathbf{Q}}X^{\boldsymbol{\sigma'}}_{l, \mathbf{Q}-\mathbf{q}'} \ , 
\end{align*}
with the exciton-magnon matrix element due to two-magnon processes reading 
\begin{equation}
G_{\boldsymbol{\sigma}\boldsymbol{\sigma}'}^{nl}(\mathbf{q})=\frac{1}{2N}[I_e F_{nl}(\boldsymbol{\beta}\cdot\mathbf{q})-I_h F_{nl}(-\boldsymbol{\alpha}\cdot\mathbf{q})]\delta_{\boldsymbol{\sigma}, \boldsymbol{\sigma}'}(-1)^{\sigma} \ , %(\delta_{\sigma_e, \sigma_h}+\delta_{\sigma_e, 1-\sigma_h}) \ , 
\end{equation}
where $\sigma=+1$ for $\sigma_e=\downarrow$ and $\sigma=-1$ for $\sigma_e=\uparrow$. Here, we note that $N=\frac{\mathcal{A}_{0}}{\mathcal{S}}$, where $\mathcal{A}_{0}$ is the magnetic unit cell area. We note that the $\delta_{\boldsymbol{\sigma}, \boldsymbol{\sigma}'}$ factor reflects that the two-magnon exciton-magnon interaction is spin-conserving.  In the considered case of CrSBr monolayers, we note that the energetic separation between spin-split states~\cite{Klein2022} and the separation between the lowest-lying exciton state to the first excited state is large and of the order of 100 meV, as evidenced by our calculations. Therefore, we find that one-magnon scattering processes are inefficient and that the two-magnon exciton-magnon Hamiltonian is reduced to an interaction between ground state excitons with a specific spin configuration. Hence, we simplify the two-magnon exciton-magnon Hamiltonian to 
\begin{equation}
H_{\mathrm{x-m}}^{(2)}=\frac{1}{\mathcal{S}}\sum_{\mathbf{Q}, \mathbf{q}, \mathbf{q}',j}G^{(\mathrm{m})}_{\mathbf{q}'}m^{\dagger}_{j,\mathbf{q}-\mathbf{q}'}m_{j,\mathbf{q}}X^{\dagger}_{\mathbf{Q}+\mathbf{q}'}X_{\mathbf{Q}} \ ,
\label{twomagham}
\end{equation}
where 
\begin{equation}
G^{(\mathrm{m})}_{\mathbf{q}}=\frac{\mathcal{A}_{0}}{2}\bigg[I_e F(\boldsymbol{\beta}\cdot\mathbf{q})-I_h F(-\boldsymbol{\alpha}\cdot\mathbf{q})\bigg] \ . 
\label{xmagcoupling}
\end{equation}

In order to get access to the magnon-induced dephasing, we start from the equation of motion for the microscopic excitonic polarization, $P^{\boldsymbol{\sigma}}_{n, \mathbf{Q}}=\langle X^{\dagger \sigma_e \sigma_h}_{n,\mathbf{Q}}\rangle$. We obtain 
\begin{equation}
\dot{P}^{\boldsymbol{\sigma}}_{n, \mathbf{Q}}=\frac{i}{\hbar}\bigg[E^{\sigma}_{n, \mathbf{Q}}P^{\boldsymbol{\sigma}}_{n, \mathbf{Q}}+\sum_{\boldsymbol{\sigma}', \mathbf{q}, l,j} M^{ln}_{\boldsymbol{\sigma}'\boldsymbol{\sigma}}(\mathbf{q})(S^{\boldsymbol{\sigma}'}_{j,l}(\mathbf{q}, \mathbf{Q})+T^{\boldsymbol{\sigma}'}_{j,l}(\mathbf{q}, \mathbf{Q}))+\sum_{l, j, \mathbf{q}', \mathbf{q}} G^{ln}_{\boldsymbol{\sigma}\boldsymbol{\sigma}}(\mathbf{q}') V^{\boldsymbol{\sigma}}_{l,j}(\mathbf{q}, \mathbf{q}', \mathbf{Q}) \bigg] \ , 
\label{eomP}
\end{equation}
upon commuting with the kinetic exciton Hamiltonian as well as the full exciton-magnon Hamiltonian. In the following, we find equations of motion for the magnon-assisted quantities $S^{\boldsymbol{\sigma}}_{j,n}(\mathbf{q}, \mathbf{Q})=\langle m^{\dagger}_{j,\mathbf{q}}X^{\dagger \boldsymbol{\sigma}}_{n, \mathbf{Q-q}}\rangle$, $T^{\boldsymbol{\sigma}}_{j,n}(\mathbf{q}, \mathbf{Q})=\langle m_{j,-\mathbf{q}}X^{\dagger \boldsymbol{\sigma}}_{n, \mathbf{Q+q}}\rangle$ and $V^{\boldsymbol{\sigma}}_{n,j}(\mathbf{q}, \mathbf{q}', \mathbf{Q})=\langle m^{\dagger}_{j,\mathbf{q}}m_{j,\mathbf{q}+\mathbf{q}'}X^{\dagger \boldsymbol{\sigma}}_{n, \mathbf{Q}+\mathbf{q}'}\rangle$.

\subsubsection{One-magnon scattering processes}
Taking into account one-magnon processes, we only have to consider the equations of motion for $S$ and $T$ in Eq.~\eqref{eomP}. We find 
\begin{align*}
\dot{S}^{\boldsymbol{\sigma}}_{j,n}(\mathbf{q}, \mathbf{Q})&=\frac{i}{\hbar}\bigg[E^{\boldsymbol{\sigma}}_{n, \mathbf{Q}-\mathbf{q}}+\hbar\Omega^{(\mathrm{m})}_{j,\mathbf{q}}\bigg]S^{\boldsymbol{\sigma}}_{j,n}(\mathbf{q}, \mathbf{Q})+\frac{i}{\hbar}\sum_{l,\boldsymbol{\sigma}'}(M^{nl}_{\boldsymbol{\sigma}\boldsymbol{\sigma}'}(\mathbf{q}))^{*}(1+n^{(\mathrm{m})}_{j,\mathbf{q}})P^{\boldsymbol{\sigma}'}_{l, \mathbf{Q}} \\ 
\dot{T}^{\boldsymbol{\sigma}}_{j,n}(\mathbf{q}, \mathbf{Q})&=\frac{i}{\hbar}\bigg[E^{\boldsymbol{\sigma}}_{n, \mathbf{Q}+\mathbf{q}}   -\hbar\Omega^{(\mathrm{m})}_{j,-\mathbf{q}} \bigg]T^{\boldsymbol{\sigma}}_{j,n}(\mathbf{q}, \mathbf{Q})+\frac{i}{\hbar}\sum_{l,\boldsymbol{\sigma}'}(M^{nl}_{\boldsymbol{\sigma}\boldsymbol{\sigma}'}(\mathbf{q}))^{*}n^{(\mathrm{m})}_{j,\mathbf{q}}P^{\boldsymbol{\sigma}'}_{l, \mathbf{Q}} \ ,
\end{align*}
where we performed the second-order Born approximation and factorized $\langle b^{\dagger} b X^{\dagger}\rangle \approx{n^{(\mathrm{m})} P}$ with $n^{(\mathrm{m})}=\langle b^{\dagger} b\rangle$ being the magnon population. Here, we also assumed the low-density limit and disregarded phonon coherences and higher-order correlations. The above equations are solved within the Markov approximation, and plugged back into the equation of motion for the polarization (Eq.~\eqref{eomP}). Neglecting off-diagonal couplings $\dot{P}^{\mu}\propto P^{\nu}$ with $\mu\neq \nu$, assuming that different exciton resonances are well separated in energy, and only considered $n=l=1s$ excitons (spin-bright and spin-dark) we find that the equation of motion for the 1\emph{s} polarization can be expressed as 
\begin{equation}
\dot{P}^{\boldsymbol{\sigma}}_{1s,\mathbf{Q}}=\frac{i}{\hbar}E^{\boldsymbol{\sigma}}_{1s, \mathbf{Q}}P^{\boldsymbol{\sigma}}_{1s,\mathbf{Q}}-\frac{\gamma^{\boldsymbol{\sigma}}_{\mathrm{m}}(\mathbf{Q})}{\hbar}|_{\mathrm{1-mag.}}P^{\boldsymbol{\sigma}}_{1s,\mathbf{Q}}
\end{equation}
where 
\begin{equation}
\gamma^{\boldsymbol{\sigma}}_{\mathrm{m}}(\mathbf{Q})|_{\mathrm{1-mag.}}=\frac{\pi}{\mathcal{S}}\sum_{\pm, \mathbf{q}, \boldsymbol{\sigma}',j}|M^{1s-1s}_{\boldsymbol{\sigma'}\boldsymbol{\sigma}}(\mathbf{q})|^2(\frac{1}{2}\pm \frac{1}{2}+n^{(\mathrm{m})}_{j,\mathbf{q}})\delta(E^{\boldsymbol{\sigma'}}_{1s, \mathbf{Q}+\mathbf{q}}-E^{\boldsymbol{\sigma}}_{1s,\mathbf{Q}}\pm \hbar\Omega^{(\mathrm{m})}_{j,\mathbf{q}})
\end{equation}
Importantly, both magnon emission and absorption processes contribute to the dephasing with spin-flipping selection rules dictated by the one-magnon exciton-magnon matrix element $M$. As the spin-split exciton bands differ substantially in energy, one-magnon processes are negligible in CrSBr at vanishing external magnetic field.

\subsubsection{Two-magnon scattering processes}
For the two-magnon contributions to the scattering process, we have to consider the two-magnon assisted quantity $V$ in Eq.~\eqref{eomP}. The equation of motion for $V$ reads 
\begin{align*}
-i\hbar\dot{V}^{\boldsymbol{\sigma}}_{n,j}(\mathbf{q}, \mathbf{q}', \mathbf{Q})&=\bigg[ E^{\boldsymbol{\sigma}}_{n,\mathbf{Q}+\mathbf{q}'}+\hbar\Omega^{(\mathrm{m})}_{j,\mathbf{q}}-\hbar\Omega^{(\mathrm{m})}_{j,\mathbf{q+q'}} \bigg]V^{\boldsymbol{\sigma}}_{n,j}(\mathbf{q}, \mathbf{q}', \mathbf{Q})+\sum_{l} n^{(\mathrm{m})}_{j,\mathbf{q}+\mathbf{q}'}P^{\boldsymbol{\sigma}}_{l, \mathbf{Q}}(1+n^{(\mathrm{m})}_{j,\mathbf{q}})G^{ln}_{\boldsymbol{\sigma}\boldsymbol{\sigma}}(\mathbf{q}')
\end{align*}
where we disregarded terms proportional to phonon coherences ($\langle b^{(\dagger)}\rangle$) and disregarded terms that scale with the exciton density upon Hartree-Fock factorization. Furthermore, we can solve the equation for $V$ within a second-order Born approximation and applying the Markov approximation. Going back to the equation of motion for $P$ then provides the magnon-induced dephasing due to two-magnon exciton-magnon scattering for 1\emph{s} excitons: 
\begin{equation}
\gamma_{\mathrm{m}}(\mathbf{Q})|_{\mathrm{2-mag.}}=\frac{\pi}{\mathcal{S}^2}\sum_{\mathbf{q}, \mathbf{q}',j}|G^{(\mathrm{m})}_{\mathbf{q}-\mathbf{q}'}|^2 n^{(\mathrm{m})}_{j,\mathbf{q}}(1+n^{(\mathrm{m})}_{j,\mathbf{q}'})\delta(E_{1s, \mathbf{Q}+\mathbf{q}-\mathbf{q}'}-E_{1s, \mathbf{Q}}+\hbar\Omega^{(\mathrm{m})}_{j,\mathbf{q}'}-\hbar\Omega^{(\mathrm{m})}_{j,\mathbf{q}} )  \ .
\end{equation}
where we omitted the spin indices since the two-magnon interaction is spin-conserving.
The dephasing can be evaluated analytically at $Q=0$ (contribution to exciton linewidth) if a parabolic dispersion for excitons and magnons is assumed. In particular, as shown in Section II, the magnon mass is substantially larger than the exciton mass, and can therefore to a good approximation be disregarded in the energy conserving delta function. In addition, the exciton form factors inside the exciton-magnon coupling $G$ (Eq.~\eqref{xmagcoupling}) decay much slower than the magnon populations as a function of momentum. Hence, these can be approximated as $F\approx 1$. Within these approximations, we find that 
\begin{align*}
\gamma_{\mathrm{m}}(0)\approx\pi\mathcal{A}_0^2\mathcal{D}^{(\mathrm{x})} \mathcal{D}^{(\mathrm{m})}|I_e-I_h|^2\frac{(k_B T)^2}{4\Delta} \ ,
\end{align*}
introducing the exciton density of states, $\mathcal{D}^{(\mathrm{x})}=\frac{\sqrt{M^{(\mathrm{x})}_{x}M^{(\mathrm{x})}_{y}}}{2\pi \hbar^2}$ as well as the magnon density of states, $\mathcal{D}^{(\mathrm{m})}=\frac{\sqrt{M^{(\mathrm{m})}_{x}M^{(\mathrm{m})}_{y}}}{2\pi \hbar^2}$ and the magnon gap $\Delta$, where it was assumed that $\Delta \ll k_B T$. Importantly, we find that the magnon-induced linewidth scales quadratically with temperature and is inversely proportional to the magnon gap. The dephasing is hence only finite if lattice anisotropies are included in the calculation, consistent with the expectation of diverging scattering rates due to parabolic dispersions and the Mermin-Wagner theorem~\cite{PhysRevB.93.155413}.

%%%%%%%%%%%%%%%%%%%%%%%%%%%%%%%%%%%%%%%%%%%%%%%%%%%%%%%%%%%%%%%%

\section{Exciton-magnon interaction in antiferromagnetic bilayers}
The exciton-magnon interaction in antiferomagnetic CrSBr bilayers was recently derived in Ref.~\cite{iakovlev2026boltzmann}. The starting-point is similar to the one discussed in the previous section, with the additional assumption that the electrons in one layer can couple to the electrons in the other layer via the interlayer electron tunneling, $t_e$. Furthermore, due to the antiferromagnetic configuration of the bilayer, one has to introduce two sets of magnon operators $a^{(\dagger)}$ and $b^{(\dagger)}$ (cf. Section II). Within second order perturbation theory (details discussed in Ref.~\cite{iakovlev2026boltzmann}), the following exciton-magnon Hamiltonian is obtained 
\begin{equation}
H^{(\mathrm{BL})}_{\mathrm{x-m}}=-\frac{\mathcal{A}_0}{4s|E^b_{X}-E^{b}_{IX}| \mathcal{S}}\sum_{\mathbf{k}, \mathbf{k}_2, \mathbf{k_1}}t^*_{e, \mathbf{k}_2-\mathbf{k}}t_{e, \mathbf{k}-\mathbf{k}_1}(a^{\dagger}_{\mathbf{k}-\mathbf{k}_2}+b_{\mathbf{k}_2-\mathbf{k}})(a_{\mathbf{k}-\mathbf{k}_1}+b^{\dagger}_{\mathbf{k}_1-\mathbf{k}})X^{\dagger}_{\mathbf{k}_2}X_{\mathbf{k}_1} \ , 
\end{equation}
dependent on the interlayer electron tunneling matrix element, $t_{e, \mathbf{k}}=t_e \sum_{\mathbf{q}}\varphi^*_{IX, \mathbf{q+\boldsymbol{\beta}\cdot k}}\varphi_{X, \mathbf{q}}$ and the overlap of intra- and interlayer 1\emph{s} exciton wave functions as well as the difference in intra- and interlayer exciton binding energies $|E^b_{X}-E^{b}_{IX}|$. Note that the magnon operators $a^{(\dagger)}$ and $b^{(\dagger)}$ do not diagonalize the free magnon Hamiltonian, instead these operators have to be expressed in the corresponding diagonal operators via the Bogoliubov transformation in Eq.~\eqref{bogolon}. In terms of these operators, the exciton-magnon Hamiltonian becomes 
\begin{equation}
H^{(\mathrm{BL})}_{\mathrm{x-m}}=-\frac{1}{\mathcal{S}}\sum_{\mathbf{p}, \mathbf{Q}, \mathbf{q}, i=1,2}V_{i,\mathbf{p}, \mathbf{q}}[(-1)^{i}(m^{\dagger}_{i,\mathbf{p}}m^{\dagger}_{i,-\mathbf{p}-\mathbf{q}}+m_{i,-\mathbf{p}}m_{i,\mathbf{p+q}})+2m^{\dagger}_{i, \mathbf{p}}m_{i, \mathbf{p+q}}]X^{\dagger}_{\mathbf{Q+q}}X_{\mathbf{Q}} \ , 
\end{equation}
where we introduce
\begin{equation}
\begin{aligned}
V_{2,\mathbf{p}, \mathbf{q}}&=\frac{\mathcal{A}_0 t^*_{e,-\mathbf{p}}t_{e, \mathbf{p+q}}(X^{++}_{-\mathbf{p}}-X^{+-}_{-\mathbf{p}})(X^{++}_{\mathbf{p+q}}-X^{+-}_{\mathbf{p+q}})}{4s|E^b_{X}-E^{b}_{IX}| } \ , \\
V_{1,\mathbf{p}, \mathbf{q}}&=\frac{\mathcal{A}_0 t^*_{e,-\mathbf{p}}t_{e, \mathbf{p+q}}(X^{--}_{-\mathbf{p}}-X^{-+}_{-\mathbf{p}})(X^{--}_{\mathbf{p+q}}-X^{-+}_{\mathbf{p+q}})}{4s|E^b_{X}-E^{b}_{IX}| } \ . 
\end{aligned}
\end{equation}
We find that the two low-energy magnon modes experience different exciton-magnon couplings via the different Bogoliubov coefficients. Furthermore, in contrast to the monolayer, in which the exciton-magnon Hamiltonian is proportional to $m^{\dagger}m$, the antiferromagnetic exciton-magnon Hamiltonian allows exciton-magnon interaction involving two creation ($(m^{\dagger})^2$) or two annihilation ($(m^2$) magnon operators, enabling two-magnon emission and absorption scattering processes, respectively. We note that the corresponding one-magnon processes in bilayers are expected to become important in the presence of an external magnetic field~\cite{iakovlev2026boltzmann}. 

In the following, we derive the corresponding magnon-induced dephasings corresponding to two-magnon emission, absorption and scattering processes. Considering the Heisenberg equation of motion for the microscopic polarization of the 1\emph{s} intralayer exciton state in a fixed layer, $P_{\mathbf{Q}}=\langle X^{\dagger}_{\mathbf{Q}}X_{\mathbf{Q}}\rangle $, we find  
\begin{equation}
\dot{P}_{\mathbf{Q}}=-\frac{i }{\mathcal{S}\hbar}\sum_{\mathbf{p,\mathbf{q}}, j=1,2}V_{j, \mathbf{p}, \mathbf{q}}[(-1)^{j}(e_{j, \mathbf{p}, \mathbf{q}, \mathbf{Q}}+a_{j, \mathbf{p}, \mathbf{q}, \mathbf{Q}})+2s_{j, \mathbf{p}, \mathbf{q}, \mathbf{Q}}]+\frac{i}{\hbar}E_{\mathbf{Q}}P_{\mathbf{Q}} \ ,
\nonumber 
\end{equation}
introducing the following magnon-assisted quantities associated with two-magnon emission, absorption, and scattering:
\begin{equation}
e_{j, \mathbf{p}, \mathbf{q}, \mathbf{Q}}=\langle m^{\dagger}_{j,\mathbf{p}}m^{\dagger}_{j,-\mathbf{p}-\mathbf{q}}X^{\dagger}_{\mathbf{Q+q}}\rangle \ , a_{j, \mathbf{p}, \mathbf{q}, \mathbf{Q}}=\langle m_{j,-\mathbf{p}}m_{j,\mathbf{p+q}}X^{\dagger}_{\mathbf{Q+q}}\rangle \ , s_{j, \mathbf{p}, \mathbf{q}, \mathbf{Q}}=\langle m^{\dagger}_{j, \mathbf{p}}m_{j, \mathbf{p+q}} X^{\dagger}_{\mathbf{Q+q}}\rangle \ . 
\nonumber 
\end{equation}
To get further, we find equations of motion for these quantities leading to 
\begin{align*}
\dot{e}_{j, \mathbf{p}, \mathbf{q}, \mathbf{Q}}&=\frac{i}{\hbar}(\hbar\Omega^{(\mathrm{m})}_{j,\mathbf{p}}+\hbar\Omega^{(\mathrm{m})}_{j,-\mathbf{p}-\mathbf{q}}+E_{\mathbf{Q+q}})e_{j,\mathbf{p}, \mathbf{q}, \mathbf{Q}}-\frac{i}{\mathcal{S}\hbar}\sum_{\mathbf{p}', \mathbf{q}', l}V_{j, \mathbf{p}', \mathbf{q}'}(-1)^{l}
\langle m_{l, -\mathbf{p}'}m_{l, \mathbf{p'+q'}}m^{\dagger}_{j, \mathbf{p}}m^{\dagger}_{j, -\mathbf{p}-\mathbf{q}}X^{\dagger}_{\mathbf{Q+q'+q}}\rangle \\
\dot{a}_{j, \mathbf{p}, \mathbf{q}, \mathbf{Q}}&=\frac{i}{\hbar}(-\hbar\Omega^{(\mathrm{m})}_{j,-\mathbf{p}}-\hbar\Omega^{(\mathrm{m})}_{j,\mathbf{p}+\mathbf{q}}+E_{\mathbf{Q+q}})a_{j,\mathbf{p}, \mathbf{q}, \mathbf{Q}}\\
& \ \ \ \ -\frac{i}{\mathcal{S}\hbar}\sum_{\mathbf{p}', \mathbf{q}', l}V_{j, \mathbf{p}', \mathbf{q}'}(-1)^{l}
\langle m^{\dagger}_{l, \mathbf{p}'}m^{\dagger}_{l, \mathbf{-p'-q'}}m_{j, -\mathbf{p}}m_{j, \mathbf{p}+\mathbf{q}}X^{\dagger}_{\mathbf{Q+q'+q}}\rangle\\
\dot{s}_{j, \mathbf{p}, \mathbf{q}, \mathbf{Q}}&=\frac{i}{\hbar}(\hbar\Omega^{(\mathrm{m})}_{j,\mathbf{p}}-\hbar\Omega^{(\mathrm{m})}_{j,\mathbf{p}+\mathbf{q}}+E_{\mathbf{Q+q}})s_{j,\mathbf{p}, \mathbf{q}, \mathbf{Q}}-\frac{2i}{\hbar\mathcal{S}}\sum_{\mathbf{p}', \mathbf{q}',l}V_{j, \mathbf{p}', \mathbf{q}'}\langle m^{\dagger}_{l, \mathbf{p}'}m_{l, \mathbf{p'+q'}}m^{\dagger}_{j, \mathbf{p}}m_{j, \mathbf{p+q}}X^{\dagger}_{\mathbf{Q+q+q'}}\rangle \ , 
\end{align*}
where we only considered the low-density limit of excitons ($\langle X^{\dagger} X\rangle\ll 1$) and excluded terms which leads to magnon coherences, i.e. terms in which the number of magnon creation and annihilation operators are different. Next, we normal-order the expressions above and use the bosonic cluster expansion for the magnon operators. In addition, we assume a homogeneous system where the equations of motion couple back to magnon occupations, $n^{(\mathrm{m})}_i=\langle m^{\dagger}_im_j\rangle\delta_{i,j}$. Then, the relevant cluster expansion reads
\begin{equation}
\langle m^{\dagger}_1 m^{\dagger}_2 m_3m_4\rangle\approx n^{(\mathrm{m})}_{1}n^{(\mathrm{m})}_2(\delta_{2,3}\delta_{1,4}+\delta_{2,4}\delta_{1,3}) \ , 
\end{equation}
neglecting higher-order correlations. We then obtain 
\begin{align*}
\dot{e}_{j, \mathbf{p}, \mathbf{q}, \mathbf{Q}}&=\frac{i}{\hbar}(\hbar\Omega^{(\mathrm{m})}_{j,\mathbf{p}}+\hbar\Omega^{(\mathrm{m})}_{j,-\mathbf{p}-\mathbf{q}}+E_{\mathbf{Q+q}})e_{j,\mathbf{p}, \mathbf{q}, \mathbf{Q}}-\frac{2iV_{j, -\mathbf{p}, -\mathbf{q}}}{\mathcal{S}\hbar}(1+n^{(\mathrm{m})}_{j,\mathbf{p}})(1+n^{(\mathrm{m})}_{j,-\mathbf{p}-\mathbf{q}})P_{\mathbf{Q}} \\
\dot{a}_{j, \mathbf{p}, \mathbf{q}, \mathbf{Q}}&=\frac{i}{\hbar}(-\hbar\Omega^{(\mathrm{m})}_{j,-\mathbf{p}}-\hbar\Omega^{(\mathrm{m})}_{j,\mathbf{p}+\mathbf{q}}+E_{\mathbf{Q+q}})a_{j,\mathbf{p}, \mathbf{q}, \mathbf{Q}}-\frac{2iV_{j, -\mathbf{p}, -\mathbf{q}}}{\mathcal{S}\hbar}(-1)^{j}n^{(\mathrm{m})}_{j, -\mathbf{p}}n^{(\mathrm{m})}_{j, \mathbf{p+q}}P_{\mathbf{Q}}\\
\dot{s}_{j, \mathbf{p}, \mathbf{q}, \mathbf{Q}}&=\frac{i}{\hbar}(\hbar\Omega^{(\mathrm{m})}_{j,\mathbf{p}}-\hbar\Omega^{(\mathrm{m})}_{j,\mathbf{p}+\mathbf{q}}+E_{\mathbf{Q+q}})s_{j,\mathbf{p}, \mathbf{q}, \mathbf{Q}}-\frac{2iV_{j, -\mathbf{p}, -\mathbf{q}}}{\hbar\mathcal{S}}n^{(\mathrm{m})}_{j,\mathbf{p+q}}(1+n^{(\mathrm{m})}_{j,\mathbf{p}})P_{\mathbf{Q}} \ .
\end{align*}
The equations above can be solved within a Markov approximation and plugged back into the equation of motion for the excitonic polarization resulting in 
\begin{equation}
\dot{P}_{\mathbf{Q}}=\frac{i}{\hbar}E_{\mathbf{Q}}P_{\mathbf{Q}}-\frac{1}{\hbar}(\gamma^{(e)}_{\mathrm{m}}(\mathbf{Q})+\gamma^{(a)}_{\mathrm{m}}(\mathbf{Q})+\gamma^{(s)}_{\mathrm{m}}(\mathbf{Q}))P_{\mathbf{Q}} \ , 
\end{equation}
introducing the magnon-induced dephasing due to magnon emission, absorption and scattering as
\begin{align*}
\gamma^{(e)}_{\mathrm{m}}(\mathbf{Q})&=\frac{2\pi}{\mathcal{S}^2}\sum_{\mathbf{p},\mathbf{q}, j}|V_{j, \mathbf{p},\mathbf{q}}|^2(1+n^{(\mathrm{m})}_{j, \mathbf{p}})(1+n^{(\mathrm{m})}_{j,-\mathbf{q}})\delta(\hbar\Omega^{(\mathrm{m})}_{j,\mathbf{p}}+\hbar\Omega^{(\mathrm{m})}_{j,-\mathbf{q}}+E_{\mathbf{Q+q-p}}-E_{\mathbf{Q}}) \ , \\
\gamma^{(a)}_{\mathrm{m}}(\mathbf{Q})&=\frac{2\pi}{\mathcal{S}^2}\sum_{\mathbf{p},\mathbf{q}, j}|V_{j, \mathbf{p},\mathbf{q}}|^2n^{(\mathrm{m})}_{j, -\mathbf{p}}n^{(\mathrm{m})}_{j, -\mathbf{q}}\delta(-\hbar\Omega^{(\mathrm{m})}_{j,-\mathbf{p}}-\hbar\Omega^{(\mathrm{m})}_{j,\mathbf{q}}+E_{\mathbf{Q+q-p}}-E_{\mathbf{Q}}) \ , \\
\gamma^{(s)}_{\mathrm{m}}(\mathbf{Q})&=\frac{4\pi}{\mathcal{S}^2} \sum_{\mathbf{p},\mathbf{q}, j}|V_{j,\mathbf{p},\mathbf{q}}|^2n^{(\mathrm{m})}_{j, \mathbf{q}}(1+n^{(\mathrm{m})}_{j, \mathbf{p}})\delta(\hbar\Omega^{(\mathrm{m})}_{j,\mathbf{p}}-\hbar\Omega^{(\mathrm{m})}_{j,\mathbf{q}}+E_{\mathbf{Q+q-p}}-E_{\mathbf{Q}}) \ , 
\end{align*}
We note that the two-magnon emission rate is obtained from the absorption rate by taking $n\leftrightarrow n+1$ and inverting the signs in the delta function. Similar to the case of the monolayer, it can be readily shown that the overall dephasing at $Q=0$ scales quadratically with temperature. In our numerical calculations, we adapt $t_{e, \mathbf{0}}=53$ meV~\cite{iakovlev2026boltzmann, Wilson2021}, but otherwise include the fully momentum-dependent exciton-magnon matrix element in the numerical evaluation of the dephasing and exciton diffusion coefficients. Furthermore, we employ the parabolic approximation of the magnon dispersion (Eq.~\eqref{parabolic}), which holds well also for bilayers, and take the same magnon gap $\Delta=0.1$ meV for both low-energy magnon branches.